\documentclass[%
 aip,
 amsmath,amssymb,
 reprint,%
]{revtex4-1}

\usepackage{graphicx}
\usepackage{dcolumn}
\usepackage{bm}

\usepackage[utf8]{inputenc}
\usepackage[T1]{fontenc}
\usepackage{mathptmx}
\usepackage{etoolbox}
\usepackage{amssymb}
\usepackage{array}

\makeatletter
\def\@email#1#2{%
 \endgroup
 \patchcmd{\titleblock@produce}
  {\frontmatter@RRAPformat}
  {\frontmatter@RRAPformat{\produce@RRAP{*#1\href{mailto:#2}{#2}}}\frontmatter@RRAPformat}
  {}{}
}%
\makeatother
\begin{document}

\preprint{AIP/123-QED}

\title{$^{174}\text{Yb}^+$-$^{113}\text{Cd}^+$ sympathetic-cooling bi-species Coulomb crystal applied to microwave frequency standard.}
\author{Y Zheng}
\homepage{These authors contributed equally.}
\affiliation{ State Key Laboratory of Precision Measurement Technology and Instruments, Key Laboratory of Photon Measurement and Control Technology of Ministry of Education, Department of Precision Instrument, Tsinghua University, Beijing 100084, China }%
\affiliation{ Department of Physics, Tsinghua University, Beijing 100084, China }
\author{H. R. Qin}
\homepage{These authors contributed equally.}
\affiliation{ State Key Laboratory of Precision Measurement Technology and Instruments, Key Laboratory of Photon Measurement and Control Technology of Ministry of Education, Department of Precision Instrument, Tsinghua University, Beijing 100084, China }%
\affiliation{ Department of Physics, Tsinghua University, Beijing 100084, China }
\author{S. N. Miao}
\affiliation{ State Key Laboratory of Precision Measurement Technology and Instruments, Key Laboratory of Photon Measurement and Control Technology of Ministry of Education, Department of Precision Instrument, Tsinghua University, Beijing 100084, China }
\author{N. C. Xin}
\affiliation{ State Key Laboratory of Precision Measurement Technology and Instruments, Key Laboratory of Photon Measurement and Control Technology of Ministry of Education, Department of Precision Instrument, Tsinghua University, Beijing 100084, China }
\author{Y. T. Chen}
\affiliation{ State Key Laboratory of Precision Measurement Technology and Instruments, Key Laboratory of Photon Measurement and Control Technology of Ministry of Education, Department of Precision Instrument, Tsinghua University, Beijing 100084, China }
\author{J. Z. Han}
\affiliation{ State Key Laboratory of Precision Measurement Technology and Instruments, Key Laboratory of Photon Measurement and Control Technology of Ministry of Education, Department of Precision Instrument, Tsinghua University, Beijing 100084, China }
\author{J. W. Zhang}%
\homepage{Electronic mail: zhangjw@tsinghua.edu.cn}
\affiliation{ State Key Laboratory of Precision Measurement Technology and Instruments, Key Laboratory of Photon Measurement and Control Technology of Ministry of Education, Department of Precision Instrument, Tsinghua University, Beijing 100084, China }
\author{L. J. Wang} 
\affiliation{ State Key Laboratory of Precision Measurement Technology and Instruments, Key Laboratory of Photon Measurement and Control Technology of Ministry of Education, Department of Precision Instrument, Tsinghua University, Beijing 100084, China }
\affiliation{ Department of Physics, Tsinghua University, Beijing 100084, China }


\begin{abstract}
We reported the realization of a $^{174}\mathrm{Yb}^+$-$^{113}\mathrm{Cd}^+$ bi-species Coulomb crystal
comprising $^{174}\mathrm{Yb}^+$ ions as coolant
and verified its potential for application as a $^{113}\mathrm{Cd}^+$ microwave frequency standard
employing sympathetic cooling.
The two species of massive ions stably trapped in a Paul trap make up this large two-component crystal.
The $^{113}\mathrm{Cd}^+$ ions are trapped in the center, 
which reduces considerably RF heating and excess micromotion to which the $^{113}\mathrm{Cd}^+$ ions are subjected.
Under this scheme, the uncertainty due to the second-order Doppler effect is reduced to $5\times 10^{-16}$,
which represents an order of magnitude improvement over 
sympathetic cooled $^{40}\mathrm{Ca}^+$-$^{113}\mathrm{Cd}^+$ crystal.
The uncertainty from the second-order Zeeman effect, 
which contributes the largest uncertainty to the microwave-ion frequency standard,
is reduced to $4\times 10^{-16}$. 
The relevant AC Stark shift uncertainty is estimated to be 
$4\times 10^{-19}$.
These results indicate using $^{174}\mathrm{Yb}^+$ as coolant ions for $^{113}\mathrm{Cd}^+$ is far superior
and confirm the feasibility of a sympathetic-cooled cadmium-ion microwave clock system 
employing a $^{174}\mathrm{Yb}^+$-$^{113}\mathrm{Cd}^+$  two-component crystal.
\end{abstract}

\maketitle

Atomic clocks have been playing an important role in both practical applications\cite{hinkley2013atomic} and
basic physics research\cite{dzuba2016strongly,wcislo2016experimental,safronova2018search}.
Microwave clocks are widely used in satellite navigation\cite{bandi2011high,mallette2010space}, 
deep space exploration\cite{prestage2007atomic,burt2021demonstration}, time synchronization\cite{piester2011remote} 
and timekeeping\cite{diddams2004standards,burt2008compensated} because of their simple 
structure and high transportability. At present, a lot of important progress has been made with microwave ion clocks 
that employ trapped $^{199}\mathrm{Hg}^+$\;
\cite{burt2021demonstration,burt2008compensated,yan2022research}, 
$^{171}\mathrm{Yb}^+$\;\cite{mulholland2019laser,xin2022laser}, $^{113}\mathrm{Cd}^+$\;\cite{zhang2014toward,miao2015high,miao2021precision} 
ions.

In past work, we have realized highly stable and accurate microwave ion clocks based on laser-cooled $^{113}\mathrm{Cd}^+$
ions\cite{zhang2014toward,miao2015high,miao2021precision}. 
However, laser-cooled ion microwave clocks require a separate cooling process, which results in dead time. 
The dead time restricts the Dick effect limit, thereby preventing 
improvements in short-term stability. In addition, the ions are not 
cooled during microwave interrogation. The temperature increase leads to second-order Doppler shift (SODS) 
and limits the linewidth and signal-to-noise ratio (SNR) of the clock signal. 
To solve the above problems, we applied the sympathetic cooling technique to $^{113}\mathrm{Cd}^+$
ion microwave clocks\cite{qin2022high,zuo2018progress,han2021toward}.

The first ion microwave clock employing sympathetic cooling was built by the Bollinger group at NBS in 1991, 
which used $^{24}\mathrm{Mg}^+$ as coolant ions to sympathetically cool $^9\mathrm{Be}^+$ ion in a Penning trap.
Their experiment showed the potential of sympathetic cooling applied to ion microwave clock.
Since 2019, our group has been devoted to research the cadmium ion microwave frequency standard 
based on sympathetic cooling. We first used $^{24}\mathrm{Mg}^+$ as coolant\cite{zuo2019direct}. 
However, the cooling laser of $^{24}\mathrm{Mg}^+$ ions is not easy to obtain 
and the reaction between $\mathrm{Mg}^+$ and $\mathrm{H_2}$ 
in the background gas reduces cooling efficiency. 
We further chose $^{40}\mathrm{Ca}^+$ to sympathetically cool $^{113}\mathrm{Cd}^+$\;\cite{han2021toward,miao2023sympathetic}.
In 2022, we realized a high-performance $^{113}\mathrm{Cd}^+$ ion microwave frequency standard using this scheme,
which was the first sympathetically-cooled ion microwave clock in a Paul trap.
$^{113}\mathrm{Cd}^+$ ions were sympathetically cooled with laser-cooled $^{40}\mathrm{Ca}^+$. 
Its short-term frequency stability reached $3.48\times10^{-13}/\sqrt{\tau}$
with frequency uncertainty of $1.5\times10^{-14}$, 
which are both better than directly laser-cooled 
$^{113}\mathrm{Cd}^+$ microwave frequency standard\cite{qin2022high}.

Although $^{113}\mathrm{Cd}^+$ microwave clock sympathetically cooled with $^{40}\mathrm{Ca}^+$  performed well, 
several limitations remained. 
First, the mass difference between the two species of ions is large, making 
the ion separation ratio relatively large, thereby limiting the efficiency of sympathetic cooling. 
Second, the mass of $^{40}\mathrm{Ca}^+$ is smaller than $^{113}\mathrm{Cd}^+$;
therefore, the latter ions are in the outer layer of the ion crystal
surrounding the $^{40}\mathrm{Ca}^+$ ions.Being farther 
from the trap central axis also results in large RF heating. 
Moreover, the distance of $^{113}\mathrm{Cd}^+$ ions from the trap center leads to excess micro-motion\cite{berkeland1998minimization}
and second-order Doppler effect\cite{miao2022second},
that again severely limit improvements in frequency accuracy and stability. 

A scheme that uses $^{174}\mathrm{Yb}^+$ to sympathetically cool $^{113}\mathrm{Cd}^+$
was incorporated into the microwave frequency standard\cite{miao2021research} to improve performance.
While seldom studied, we demonstrated the viability of $^{174}\mathrm{Yb}^+$ sympathetically-cooling $^{113}\mathrm{Cd}^+$ 
microwave frequency standard system 
and realized a large number of sympathetically cooled $^{113}\mathrm{Cd}^+$ with laser-cooled $^{174}\mathrm{Yb}^+$.
At present, the frequency accuracy of the ion microwave clock is mainly restricted by 
the second-order Doppler frequency shift(SODS) and the second-order Zeeman frequency shift(SOZS). 
Under this scheme, the $^{113}\mathrm{Cd}^+$ ion temperature is as low as $10^2\text{mK}$ and the excess micro-motion
is greatly suppressed.
The uncertainty associated with the SODS is reduced to $5\times 10^{-16}$,
which is advanced by an order of magnitude compared with that of $^{40}\mathrm{Ca}^+$-$^{113}\mathrm{Cd}^+$ sympathetic cooling.
The uncertainty related to SOZS is reduced to $4\times 10^{-16}$.
Our research shows that sympathetic-cooled mixed-species Coulomb crystal of $^{174}\mathrm{Yb}^+$-$^{113}\mathrm{Cd}^+$  
is an effective experimental method that improves the performance of the $^{113}\mathrm{Cd}^+$ ion microwave frequency standard.\\
\indent The ion trap we used has been described in more detail in our previous work\cite{miao2021precision}. 
The ground state hyperfine splitting frequency of $^{113}\mathrm{Cd}^+$ is 15.2 GHz, 
ranking second only to $^{199}\mathrm{Hg}^+$ among all working energy levels of atomic microwave clocks.
Because the hyperfine splitting frequency of the $^2P_{3/2}$ energy level is only 800 MHz, 
pumping and detection can be realized by a single laser 
with acousto-optic modulators. 
Therefore, the $^{113}\mathrm{Cd}^+$ frequency standard has great potential for high performance and miniaturization.
The natural abundance of $^{174}\mathrm{Yb}$ is 31.83\%, which is the highest among the seven stable Yb isotopes. 
Moreover, the cooling laser $^{174}\mathrm{Yb}^+$ requires is easily available. Compared with $^{40}\mathrm{Ca}^+$, 
$^{174}\mathrm{Yb}^+$ is not only closer in mass but also heavier than $^{113}\mathrm{Cd}^+$. 
These characteristics indicate that $^{174}\mathrm{Yb}^+$ is well suited as a coolant ion for $^{113}\mathrm{Cd}^+$.

\begin{figure}
 \centering
\includegraphics[width=0.45\textwidth]{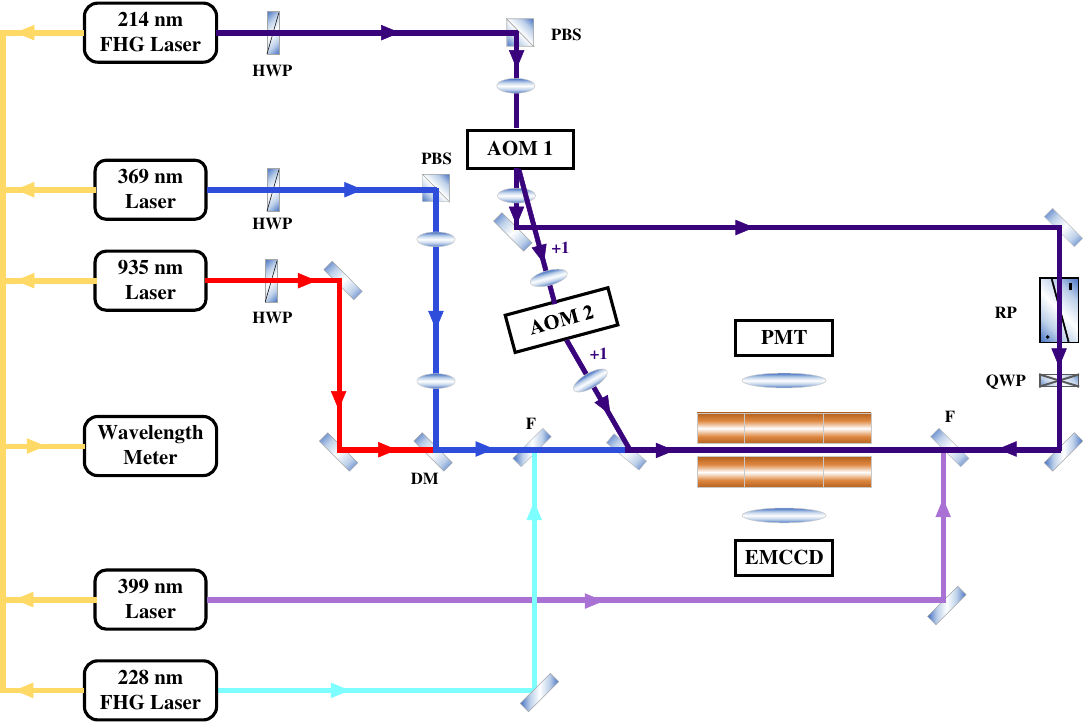}
 \caption{Schematic diagram of the entire optic system designed. HWP:
half-wave plate; PBS: polarization beam splitter; 
RP: Rochon polarizer; QWP, quarter-wave plate;
DM: dichromatic mirror;
F: mirror installed on flipper; 
AOM: acousto-optic modulator. }
 \label{fig1}
\end{figure}

The entire optical system we designed (Fig.~\ref{fig1}), 
incorporates the $^{113}\mathrm{Cd}^+$ component of a previous design.\cite{qin2022high}. 
The 399 nm laser beam is used to transition $^{174}\mathrm{Yb}$ atom from the ground state $6s^2\,{^1S_0}$ 
to the first excited state $6s6p\,^1P_1$.
Because there is an isotopic shift of approximately 500 MHz between $^{174}\mathrm{Yb}$ and other isotopes\cite{guoliming}, 
this photoionization selectively ionizes the $^{174}\mathrm{Yb}$ atoms.
For Doppler cooling and repumping of the $^{174}\mathrm{Yb}^+$,
the 369 nm laser beam ($6s\,{^2S_{1/2}}\rightarrow6p\,{^2P_{1/2}}$) is combined with 
the 935 nm laser beam ($^2D_{3/2}\rightarrow ^3[3/2]_{1/2}$) by the dichroic mirror.
Photoionization is enabled by turning on the cooling laser with the wavelength of 369 nm 
while opening the laser with the wavelength of 399 nm to excite the $^{174}\mathrm{Yb}$ atoms.
\\
\begin{figure}[thbp]
\centering
\includegraphics[width=0.45\textwidth]{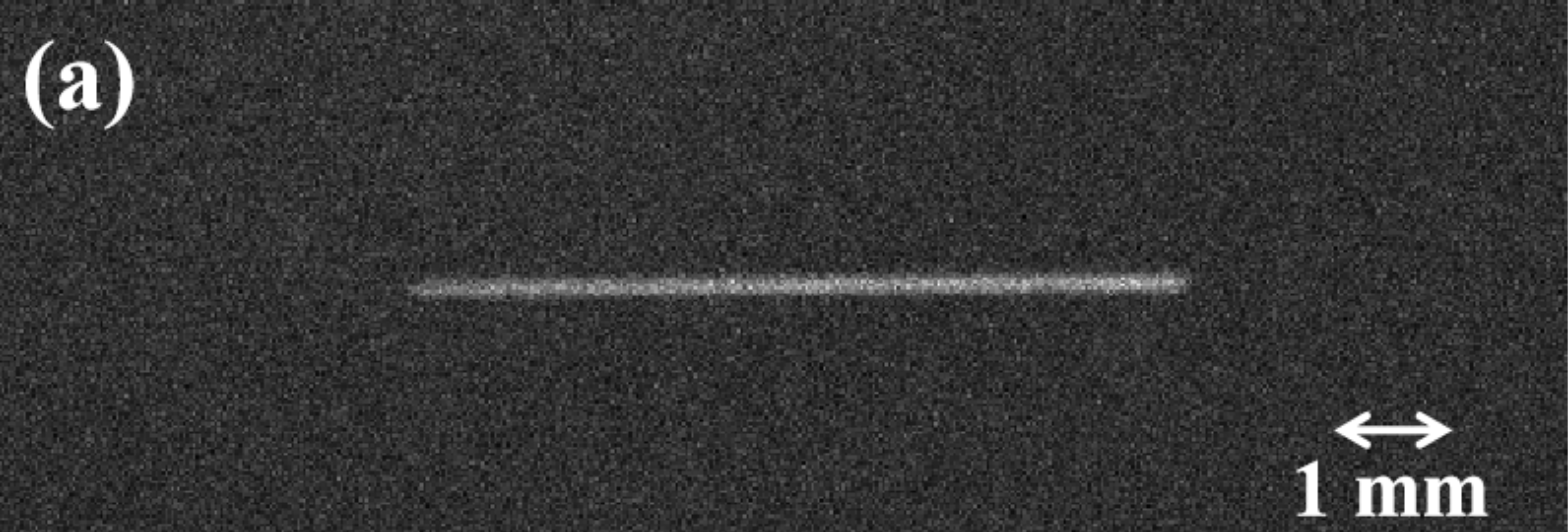}
\includegraphics[width=0.45\textwidth]{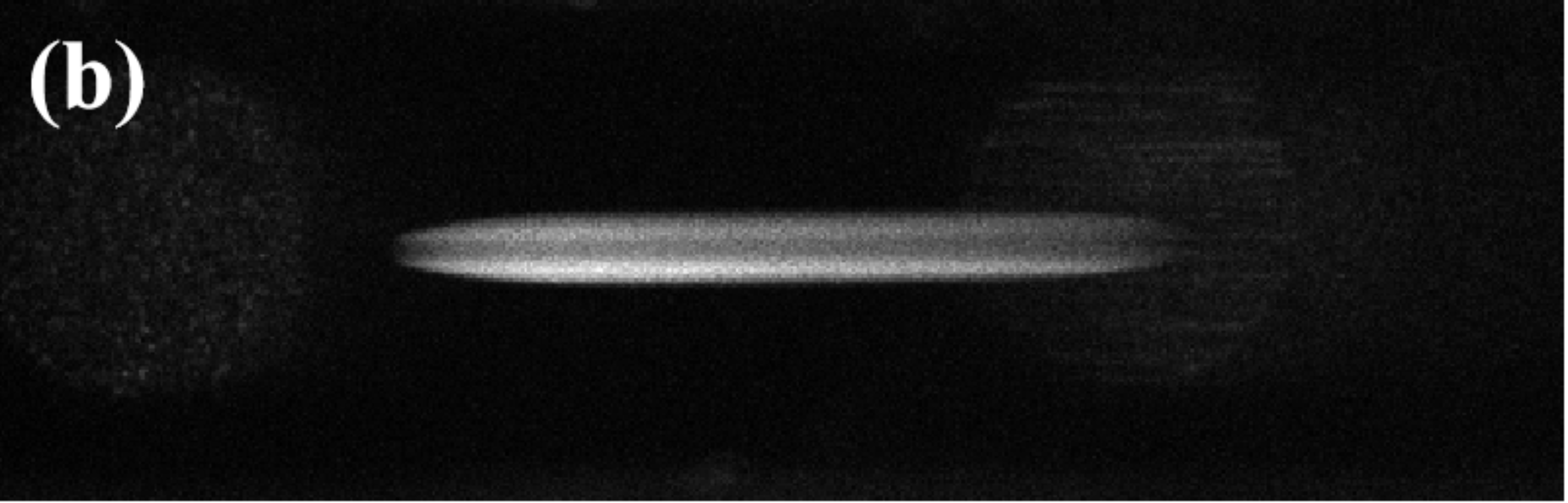}
\includegraphics[width=0.45\textwidth]{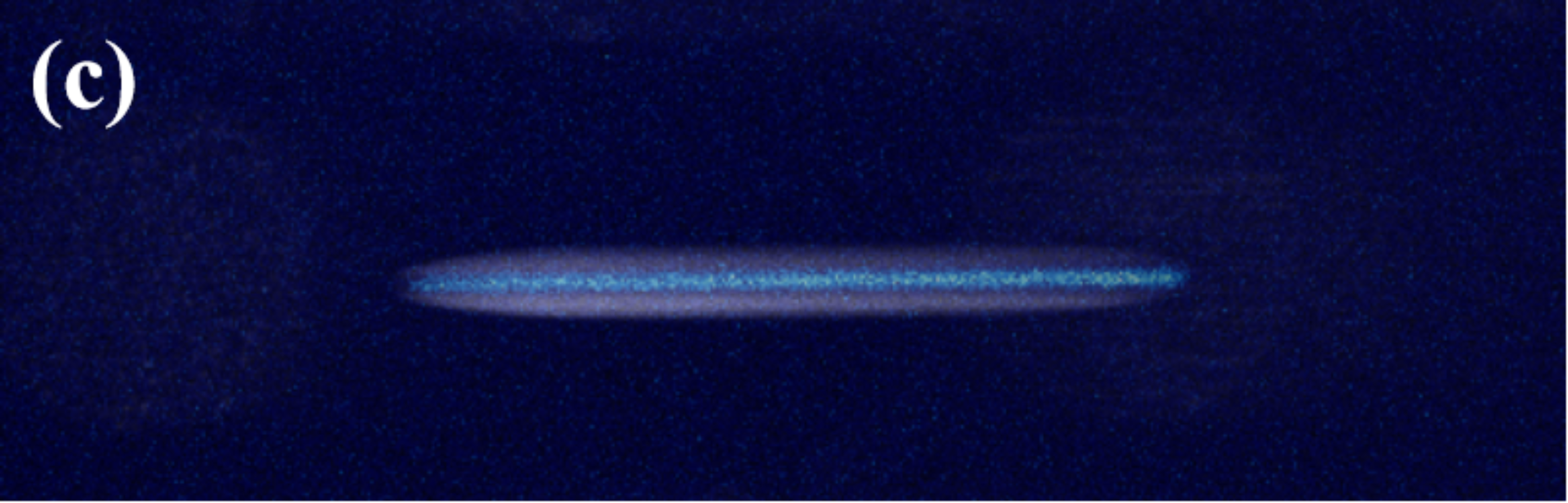}
\caption{Typical image of $^{174}\mathrm{Yb}^+$-$^{113}\mathrm{Cd}^+$ two-component ion crystal 
 with $U_{RF}=241\mathrm{V}$. 
 (a) Image of $^{113}\mathrm{Cd}^+$ under 214.5 nm wavelength filter.
 (b) Image of $^{174}\mathrm{Yb}^+$ under 397 nm wavelength filter.
(c) Image combined of two images taken by EMCCD under different wavelength filters.}
\label{fig2}
\end{figure}
\indent In the $^{174}\mathrm{Yb}^+$-$^{113}\mathrm{Cd}^+$ sympathetic cooling stage,
we loaded and cooled the $^{174}\mathrm{Yb}^+$ ions to form the crystal first and subsequently loaded the $^{113}\mathrm{Cd}^+$ ions. 
Then $^{174}\mathrm{Yb}^+$ were heated. We repeated the adjustments to the amplitude of RF voltage
and scanned the frequency of the cooling laser (369 nm). 
The detection by the photomultiplier tube (PMT) of a variation in the fluorescence signal of $^{113}\mathrm{Cd}^+$ 
indicates the two species of ions have formed a two-component Coulomb crystal. 

A typical image captured by the electron-multiplying charge-coupled device (EMCCD) of the two-component ion crystal (Fig.~\ref{fig2})
depicts a hollow structure of $^{174}\mathrm{Yb}^+$ ions and an ellipsoid of $^{113}\mathrm{Cd}^+$ ions located in the center, 
which is in line with our expectation in using $^{174}\mathrm{Yb}^+$ as a coolant. 
To estimate the number of ions, we need to determine the separation ratio 
and density of the two species of ions. 
The separation ratio is calculated using
\cite{hornekaer2000single,hornekaer2001structural,o1981centrifugal,wineland1987ion} :
\begin{equation}
\frac{r_{\mathrm{Yb}^+}}{r_{\mathrm{Cd}^+}}=\sqrt{\frac{M_{\mathrm{Yb}^+}}{M_{\mathrm{Cd}^+}}} ,
\end{equation}
where $r_{\mathrm{Yb}^+}$ denotes the inner radius of the $^{174}\mathrm{Yb}^+$ crystal shell, 
$r_{\mathrm{Cd}^+}$ the outer radius of the $^{113}\mathrm{Cd}^+$ crystal.
We experimentally measured the separation ratio by analyzing the EMCCD images.
The result is $r_{\mathrm{Yb}^+}/r_{\mathrm{Cd}^+}=1.25(3)$, 
which agrees with the theoretical value of 1.24.
Using the zero-temperature charged-liquid model
\cite{wineland1987ion,hornekaer2001structural}, the ion density is estimated to be:
\begin{equation}
 n=\frac{\varepsilon_{0} U_\mathrm{R F}^{2}}{M \Omega^{2} r_{0}^{4}},
\end{equation}
where $\varepsilon_{0}$ denotes the permittivity of vacuum, $U_\mathrm{RF}$ the RF amplitude,
$M$ the trapped ion mass, $r_0=6.2\;\text{mm}$ the radial distance from the trap axis to the electrodes,
and $\Omega$ the trap driving frequency. 
In our experiment setup, $U_\mathrm{RF}=241\;\mathrm{V}$ and $\Omega=2\pi\times1.994\;\mathrm{MHz}$.
The ion densities of $^{174}\mathrm{Yb}^+$ and $^{113}\mathrm{Cd}^+$ were thus estimated to be
$7.67\times10^3\mathrm{mm}^{-3}$ and $1.18\times10^4\mathrm{mm}^{-3}$, respectively. 
\begin{figure}[tbhp]
\includegraphics[width=0.45\textwidth]{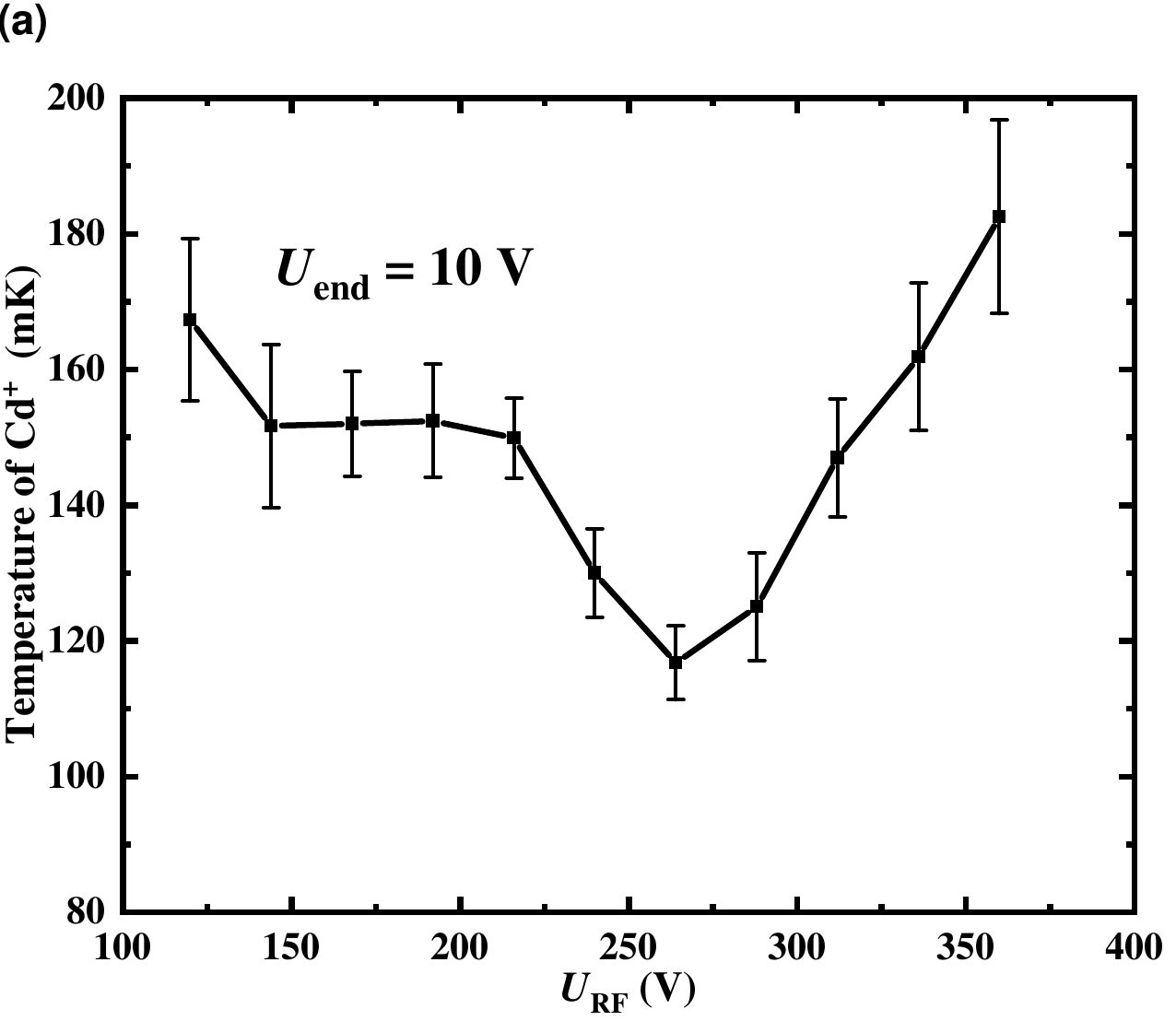}
\includegraphics[width=0.45\textwidth]{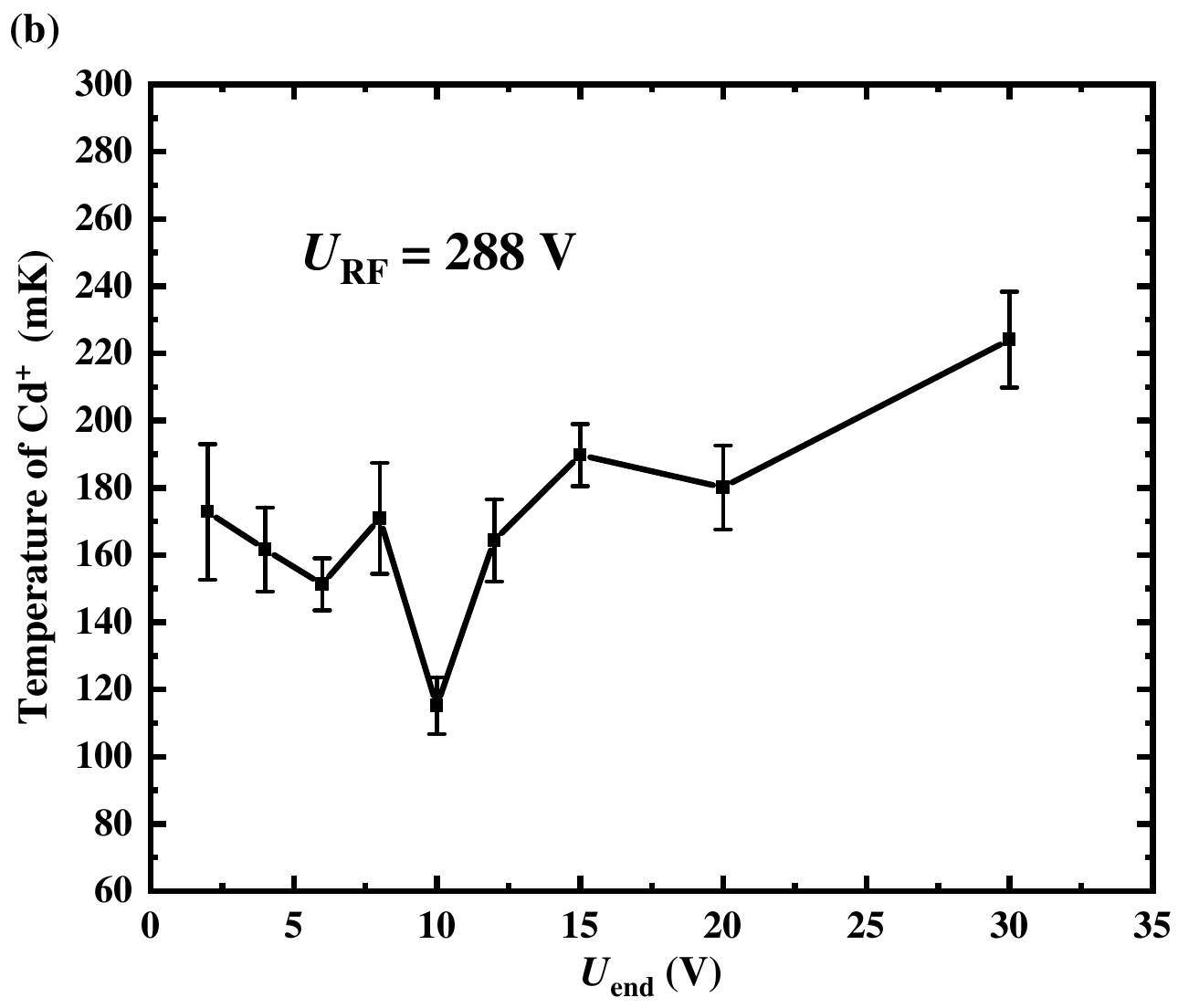}
\caption{Influence of electrical parameters on the temperature of sympathetically cooled $^{113}\mathrm{Cd}^+$.
(a) Temperature at different $U_\mathrm{RF}$ with $U_\mathrm{end}=10\mathrm{V}$.
(b) Temperature at different $U_\mathrm{end}$ with $U_\mathrm{RF}=288\mathrm{V}$.}
\label{fig3}
\end{figure}
Typical population numbers of the $^{174}\mathrm{Yb}^+$ and $^{113}\mathrm{Cd}^+$ ions
in the two-component crystal are $N_\mathrm{{Yb}^{+}}=3.3(4)\times10^{3}$ and 
$N_\mathrm{{Cd}^{+}}=6.9(3)\times10^{3}$,
which is more than that in a $^{40}\mathrm{Ca}^+$-$^{113}\mathrm{Cd}^+$ 
sympathetic cooled crystal.

The SNR of microwave frequency standard depends mainly on the number of trapped ions.
While $^{113}\mathrm{Cd}^+$ ions were trapped in the outer shell of the $^{40}\mathrm{Ca}^+$-$^{113}\mathrm{Cd}^+$ two-component Coulomb crystal,
RF heating and second-order Doppler effect prevent increasing the number of ions.
Trapping $^{113}\mathrm{Cd}^+$ ions in the center avoids the problem
and by increasing the number of $^{113}\mathrm{Cd}^+$ ions we could improve the SNR of the system.

After trapping the mixed-species Coulomb crystal, 
the effect of RF voltage and endcap voltage($U_\mathrm{end}$) on the temperature of the $^{113}\mathrm{Cd}^+$ were explored.
The temperature is calculated by 
measuring the Doppler broadening of the fluorescence spectrum\cite{larson1986sympathetic,bollinger1984strongly}.
Fluorescence line of ions usually follows a Voigt line, 
a convolution of the natural linewidth and Gaussian width of the Doppler broadening.
After obtaining the Gaussian linewidth, the ion temperature is calculated using
\begin{equation}
 T=\frac{M c^2}{8\ln2\times{k_{B}}}\left(\frac{\mathrm{\nu}_G}{\mathrm{\nu}_c}\right)^2,
\end{equation}
where $c$ denotes the speed of light, $k_{B}$ the Boltzmann constant,
$\mathrm{\nu}_G$ the fitted Gaussian linewidth and 
$\mathrm{\nu}_c$ the resonance frequency of the $D_2$ line of $^{113}\mathrm{Cd}^+$.
Fig.~\ref{fig3} (a) reveals the dependence of the temperature of sympathetically-cooled $^{113}\mathrm{Cd}^+$
on $U_\mathrm{RF}$. 
As RF voltage increases, the radial size of the $^{113}\mathrm{Cd}^+$ ion crystal is compressed, 
leading to less RF heating but resulting in an increase in micro-motion energy.
Thus, there is an optimal value for RF voltage of approximately $U_\mathrm{RF}=264\;\mathrm{V}$. 
Similarly, increasing $U_\mathrm{end}$ increases the ion radial size.
However, if $U_\mathrm{end}$ is too low, the ion trap becomes unstable.
The optimal $U_\mathrm{end}$ is approximately $10\;\mathrm{V}$ as shown in Fig.~\ref{fig3} (b).

\begin{figure}[!t]
 \centering
\includegraphics[width=0.45\textwidth]{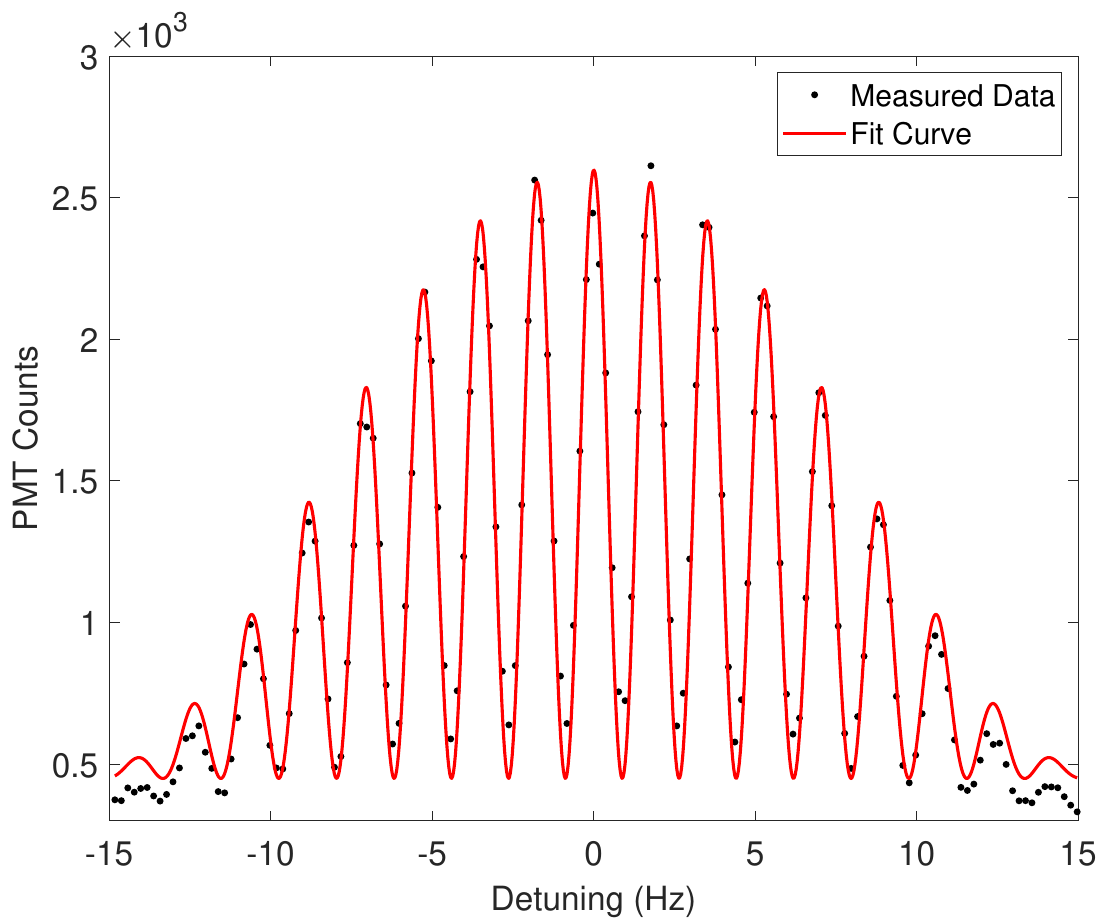}
 \caption{Typical Ramsey fringe of the clock transition (15.2 GHz) with a free evolution time of 500 ms. }
\label{fig4}
\end{figure}

In this situation, a preliminary-acquired Ramsey fringe of the clock transition (Fig.~\ref{fig4}) is obtained with
a free evolution time of 500 ms. We will enhance the SNR by
optimizing the electric parameters and the population number ratio
to realize a sympathetically-cooled ion microwave frequency standard exhibiting high performance.
The expected short-term and long-term frequency stabilities are $2 \times10^{-13}/\sqrt{\tau}$
and $5\times10^{-15}@10000s$.

The main uncertainties of frequency shifts in the $^{174}\mathrm{Yb}^+$-$^{113}\mathrm{Cd}^+$ sympathetic cooling system 
were carefully evaluated. The SOZS, which is the main source of systematic uncertainty for an ion microwave frequency standard
is given by
\begin{equation}
\begin{aligned}
\frac{{\delta}\nu_{\text{SOZFS}}}{\nu_0}& =\frac{\left(g_J-g_I\right)^2\mu_B^2}{2h^2\nu_0^2}B^2\\
\end{aligned}
\end{equation}
where $\mu_B$ denotes Bohr magneton, $B$ the magnetic field intensity, $h$ Planck constant,
$\nu_0$ the transition frequency 
of $^{113}\mathrm{Cd}^+$ between states $|^2S_{1/2}, F=0, m_F=0\rangle$ and $|^2S_{1/2}, F=1, m_F=0\rangle$
at zero magnetic field, 
and for which the values for the electronic and nuclear Landé g-factors\cite{han2022determination,spence1972optical}
are ${g}_{J}=2.002 291(4)$ and ${g}_{I}=0.622 300 9(9)\times10^{-3}$, respectively.
Because the $^{113}\mathrm{Cd}^+$ ions were trapped in the center, 
where the magnetic gradient perpendicular to the quantization axis ($\boldsymbol{e_z}$) is smaller,
the magnetic field required to provide the quantization axis for clock ions was reduced
while measuring the clock transition signal of the $^{113}\mathrm{Cd}^+$ ions 
sympathetically cooled via the $^{174}\mathrm{Yb}^+$ ions
with the magnetic fields along $\boldsymbol{e_x}$ and $\boldsymbol{e_y}$ well compensated.
While collecting the Ramsey signal of the microwave clock transition,
the static magnetic field is measured to be 
648.1 nT,
an order of magnitude smaller than before\cite{qin2022high}.
The fluctuation of the magnitude field under our high-performance magnetic shielding can
be weakened to 0.18 nT \cite{xin2022laser}.
Thus, SOZS is estimated to be $7.133(4)\times10^{-13}$.
Compared with the previous generation of sympathetically-cooled microwave clock, 
the absolute value of SOZS is reduced by more than two orders of magnitude 
and the uncertainty is raised by more than two orders of magnitude.

\begin{figure}[thbp]
\includegraphics[width=0.45\textwidth]{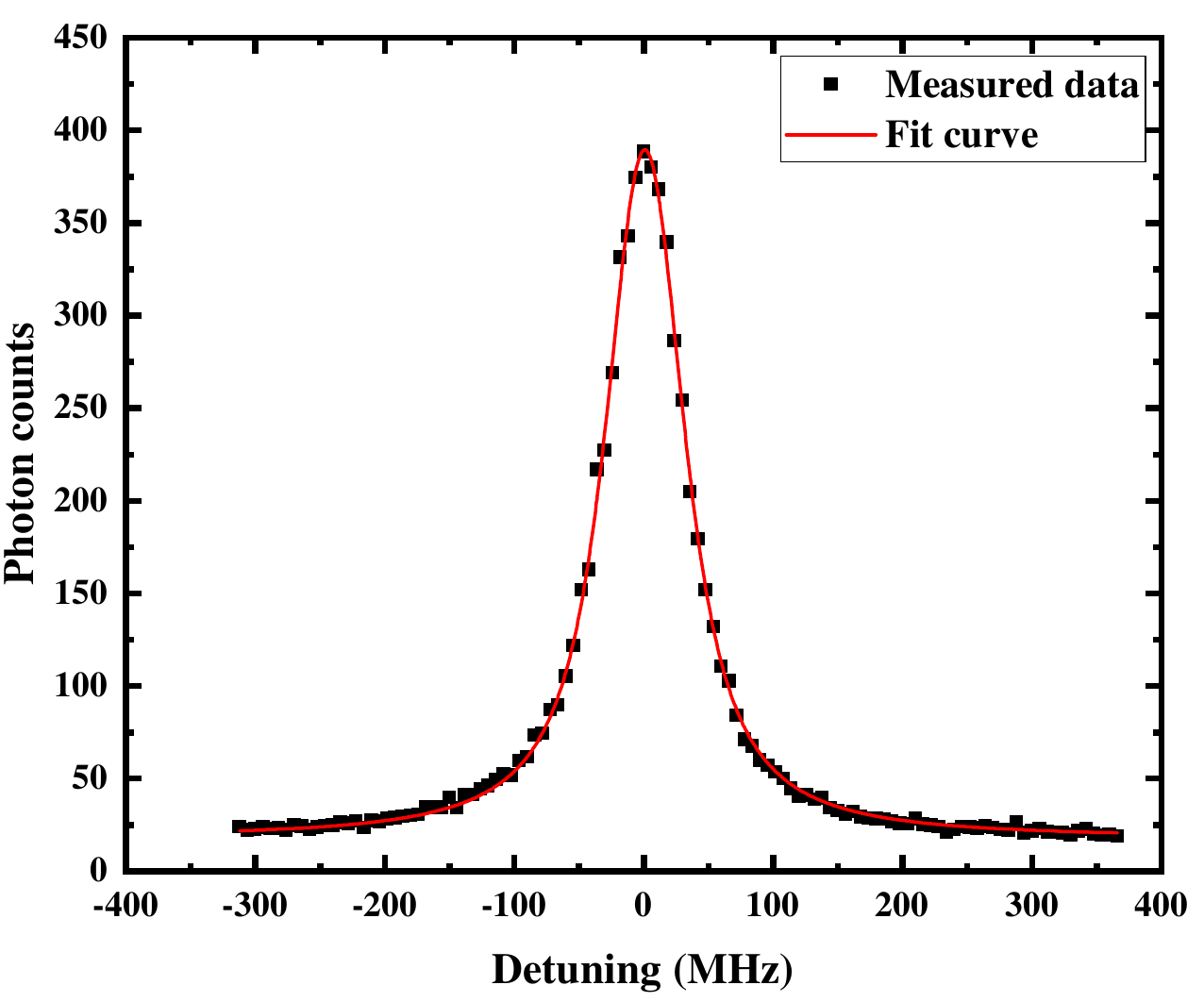}
\caption{Typical fluorescence line of temperature measurement of sympathetically-cooled $^{113}\mathrm{Cd}^+$.}
\label{fig5}
\end{figure}

One important reason why we introduced $^{174}\mathrm{Yb}^+$ ions as coolant is to 
reduce SODS. Secular motion, micro-motion and excess micro-motion through deviations from the central axis of the trap
contribute to SODS. The first two contributions correlate with the secular temperature of the ions, 
and the last is determined by the position of the $^{113}\mathrm{Cd}^+$ ion crystal\cite{miao2022second}.
In the experiment, the typical temperature measurement result of sympathetically-cooled $^{113}\mathrm{Cd}^+$ is shown in Fig.~\ref{fig5}.
The corresponding ion temperature is $100(5)\mathrm{mK}$,
much smaller than that of laser-cooled $^{113}\mathrm{Cd}^+$
(654 mK)\cite{miao2021precision}.
The secular motion and micro-motion of the trapped ions are of the same magnitude in all directions
\cite{berkeland1998minimization}.
Therefore, the reduction in the axial temperature we measured 
is crucial for the further development of  microwave ion clocks.
The formula of the SODS caused contributed by excess micro-motion is\cite{miao2022second}:
\begin{equation}
 \begin{aligned}
\frac{\delta\nu_{\text{SODS-exmm}}}{\nu_0}& =-\frac{1}{16}\frac{q^2\Omega^2u^2}{c^2} \\
q& =\frac{2Q U_{\mathrm{RF}}}{M\Omega^{2}r_{0}^{2}}
\end{aligned}
\end{equation}
where $Q$ denotes the charge of $^{113}\mathrm{Cd}^+$ and $u$ the distance of ions
from the central axis.
For large ion crystals, it is inevitable that ions deviate from the central axis of the Paul trap.
This problem is particularly prominent when $^{113}\mathrm{Cd}^+$ ions were cooled by $^{40}\mathrm{Ca}^+$ ions
and located in the outer shell\cite{qin2022high}. Our method mitigates this effect which is why we consider 
$^{174}\mathrm{Yb}^+$-$^{113}\mathrm{Cd}^+$ sympathetic cooling suitable
for microwave ion frequency standard.
Calculating distances from the EMCCD images,
the SODS involved with excess micro-motion is estimated to be $-7.8(5)\times 10^{-15}$,
which is three times smaller than that of $^{40}\mathrm{Ca}^+$-$^{113}\mathrm{Cd}^+$ sympathetic cooling. 
Moreover, the uncertainty is six times smaller than before\cite{qin2022high}.
The SODS contributed by secular motion and micro-motion is calculated to be $-3.7(2)\times 10^{-16}$.
Finally, the total SODS is estimated to be $-8.1(5)\times 10^{-15}$.

The Stark shift generated by the additional static electric field 
can be described by the following formula\cite{berkeland1998minimization}:
\begin{equation}
\frac{\delta\nu_{\mathrm{DC-S}}}{\nu_0}=-\frac{2\sigma_S}{\nu_0}\left(\frac{m\Omega c}{Q}\right)^2\frac{\delta\nu_{\mathrm{SODS-exmm}}}{\nu_0},
\end{equation}
where $\sigma_S$ denotes the static Stark shift coefficient.
This term is proportional to the SODS produced by excess micro-motion and therefore is 
reduced to $7.9(5)\times10^{-17}$.

\begin{table*}
\caption{\label{tab:table1}Comparison of the fundamental fractional systematic uncertainties 
for frequency shifts among the laser-cooling, $^{40}\mathrm{Ca}^+$-$^{113}\mathrm{Cd}^+$ sympathetic cooling
and $^{174}\mathrm{Yb}^+$-$^{113}\mathrm{Cd}^+$ sympathetic cooling microwave frequency standards.  }
\begin{ruledtabular}
\begin{tabular}{cccc}
Item & Laser-cooling & $^{40}\mathrm{Ca}^+$-$^{113}\mathrm{Cd}^+$ & $^{174}\mathrm{Yb}^+$-$^{113}\mathrm{Cd}^+$\\
\hline
\\
SOZS uncertainty & $3\times10^{-15}$ & $1.1\times10^{-14}$ & $4\times10^{-16}$\\
SODS uncertainty & $3.6\times10^{-15}$ & $3\times10^{-15}$ & $5\times10^{-16}$\\
AC Stark shift (laser) & ~0 & $5.4(5)\times10^{-17}$ & $7.98(4)\times10^{-17}$\\
Additional Stark shift & Not Mentioned & $2.3(3)\times10^{-16}$ & $7.9(5)\times10^{-17}$
\end{tabular}
\end{ruledtabular}
\end{table*}

Additional ac Stark(light) shifts introduced by the cooling(369 nm) and 
repumping(935nm) laser beams of $^{174}\mathrm{Yb}^+$ were evaluated. 
The intensities of the 369 nm and 935 nm beams were $0.264(3)\,\mathrm{mW/mm^2}$ and $2.123(6)\,\mathrm{mW/mm^2}$, respectively.
yielding the light shifts of $1.77(2)\times10^{-17}$ and $6.21(2)\times10^{-17}$, respectively,
which can be nearly negligible for microwave frequency standard.

In conclusion, we performed a study of $^{174}\mathrm{Yb}^+$-$^{113}\mathrm{Cd}^+$ sympathetic cooling
and proved its key advantages in a microwave ion frequency standard.
Compared with laser cooling, sympathetic cooling overcomes the rising temperature of ions during interrogation,
reduces the dead time, and greatly prolongs the free evolution time in closed-loop locking. 
Compared with using $^{40}\mathrm{Ca}^+$ to sympathetically cool $^{113}\mathrm{Cd}^+$,
the mass difference between $^{174}\mathrm{Yb}^+$ and $^{113}\mathrm{Cd}^+$ is smaller,
which is advantageous in promoting cooling efficiency and extending ion-loss time.
Table~\ref{tab:table1} lists the fundamental fractional systematic uncertainties for frequency shifts
among the three schemes\cite{miao2021precision,qin2022high}.
Under $^{174}\mathrm{Yb}^+$-$^{113}\mathrm{Cd}^+$ sympathetic cooling,
the SODS uncertainty was reduced to $5\times 10^{-16}$ 
because excess micro-motion was substantially suppressed. 
the SOZS and its uncertainty were also considerably reduced to $7.133(4)\times 10^{-13}$.
In addition, the introduced Stark shift from the static electric field 
and its uncertainty were estimated to $7.9(5)\times10^{-17}$,
which is superior by nearly an order of magnitude to our prior work\cite{qin2022high}.
The light frequency shift was evaluated to be $7.98(4)\times10^{-17}$.
These results show that $^{174}\mathrm{Yb}^+$-$^{113}\mathrm{Cd}^+$ sympathetic cooling applied to
a microwave frequency standard promises to attain high accuracy  of $10^{-16}$ and high stability.

In the future, we will study the thermodynamic properties of the $^{174}\mathrm{Yb}^+$-$^{113}\mathrm{Cd}^+$ 
ion crystal at low temperatures based on sympathetic cooling 
in combination with molecular dynamics simulations and experiments 
to  elicit further advantages for the microwave frequency standard.

The $^{174}\mathrm{Yb}^+$-$^{113}\mathrm{Cd}^+$ crystallization results enrich the research
on large two-component ion crystals, which will be meaningful for research on
structures of Coulomb crystals\cite{richerme20162d,d2021radial}, 
high-precision measurements of isotope shifts\cite{gebert2015precision},
the dynamics of an ion or small ion crystal\cite{guggemos2015sympathetic},
radioactive ions\cite{groot2019trapping} and quantum simulations\cite{raghunandan2020initialization}
based on the sympathetic cooling technique.

\begin{acknowledgments}
This work is supported by National Key Research and Development Program of China (
2021YFA1402100), National Natural Science 
Foundation of China (12073015) and the Science and Technology on Metrology and Calibration Laboratory (Grant No. JLKG2022001A002).
\end{acknowledgments}
\section*{Data Availability Statement}
Data supporting the findings of this study are available upon reasonable request from the corresponding author.

\nocite{*}
\bibliography{aipsamp}

\providecommand{\noopsort}[1]{}\providecommand{\singleletter}[1]{#1}%
\begin{thebibliography}{40}%
\makeatletter
\providecommand \@ifxundefined [1]{%
 \@ifx{#1\undefined}
}%
\providecommand \@ifnum [1]{%
 \ifnum #1\expandafter \@firstoftwo
 \else \expandafter \@secondoftwo
 \fi
}%
\providecommand \@ifx [1]{%
 \ifx #1\expandafter \@firstoftwo
 \else \expandafter \@secondoftwo
 \fi
}%
\providecommand \natexlab [1]{#1}%
\providecommand \enquote  [1]{``#1''}%
\providecommand \bibnamefont  [1]{#1}%
\providecommand \bibfnamefont [1]{#1}%
\providecommand \citenamefont [1]{#1}%
\providecommand \href@noop [0]{\@secondoftwo}%
\providecommand \href [0]{\begingroup \@sanitize@url \@href}%
\providecommand \@href[1]{\@@startlink{#1}\@@href}%
\providecommand \@@href[1]{\endgroup#1\@@endlink}%
\providecommand \@sanitize@url [0]{\catcode `\\12\catcode `\$12\catcode
  `\&12\catcode `\#12\catcode `\^12\catcode `\_12\catcode `\%12\relax}%
\providecommand \@@startlink[1]{}%
\providecommand \@@endlink[0]{}%
\providecommand \url  [0]{\begingroup\@sanitize@url \@url }%
\providecommand \@url [1]{\endgroup\@href {#1}{\urlprefix }}%
\providecommand \urlprefix  [0]{URL }%
\providecommand \Eprint [0]{\href }%
\providecommand \doibase [0]{http://dx.doi.org/}%
\providecommand \selectlanguage [0]{\@gobble}%
\providecommand \bibinfo  [0]{\@secondoftwo}%
\providecommand \bibfield  [0]{\@secondoftwo}%
\providecommand \translation [1]{[#1]}%
\providecommand \BibitemOpen [0]{}%
\providecommand \bibitemStop [0]{}%
\providecommand \bibitemNoStop [0]{.\EOS\space}%
\providecommand \EOS [0]{\spacefactor3000\relax}%
\providecommand \BibitemShut  [1]{\csname bibitem#1\endcsname}%
\let\auto@bib@innerbib\@empty
\bibitem [{\citenamefont {Hinkley}\ \emph {et~al.}(2013)\citenamefont
  {Hinkley}, \citenamefont {Sherman}, \citenamefont {Phillips}, \citenamefont
  {Schioppo}, \citenamefont {Lemke}, \citenamefont {Beloy}, \citenamefont
  {Pizzocaro}, \citenamefont {Oates},\ and\ \citenamefont
  {Ludlow}}]{hinkley2013atomic}%
  \BibitemOpen
  \bibfield  {author} {\bibinfo {author} {\bibfnamefont {N.}~\bibnamefont
  {Hinkley}}, \bibinfo {author} {\bibfnamefont {J.~A.}\ \bibnamefont
  {Sherman}}, \bibinfo {author} {\bibfnamefont {N.~B.}\ \bibnamefont
  {Phillips}}, \bibinfo {author} {\bibfnamefont {M.}~\bibnamefont {Schioppo}},
  \bibinfo {author} {\bibfnamefont {N.~D.}\ \bibnamefont {Lemke}}, \bibinfo
  {author} {\bibfnamefont {K.}~\bibnamefont {Beloy}}, \bibinfo {author}
  {\bibfnamefont {M.}~\bibnamefont {Pizzocaro}}, \bibinfo {author}
  {\bibfnamefont {C.~W.}\ \bibnamefont {Oates}}, \ and\ \bibinfo {author}
  {\bibfnamefont {A.~D.}\ \bibnamefont {Ludlow}},\ }\bibfield  {title}
  {\enquote {\bibinfo {title} {An atomic clock with 10--18 instability},}\
  }\href@noop {} {\bibfield  {journal} {\bibinfo  {journal} {Science}\ }\textbf
  {\bibinfo {volume} {341}},\ \bibinfo {pages} {1215--1218} (\bibinfo {year}
  {2013})}\BibitemShut {NoStop}%
\bibitem [{\citenamefont {Dzuba}\ \emph {et~al.}(2016)\citenamefont {Dzuba},
  \citenamefont {Flambaum}, \citenamefont {Safronova}, \citenamefont {Porsev},
  \citenamefont {Pruttivarasin}, \citenamefont {Hohensee},\ and\ \citenamefont
  {H{\"a}ffner}}]{dzuba2016strongly}%
  \BibitemOpen
  \bibfield  {author} {\bibinfo {author} {\bibfnamefont {V.}~\bibnamefont
  {Dzuba}}, \bibinfo {author} {\bibfnamefont {V.}~\bibnamefont {Flambaum}},
  \bibinfo {author} {\bibfnamefont {M.}~\bibnamefont {Safronova}}, \bibinfo
  {author} {\bibfnamefont {S.}~\bibnamefont {Porsev}}, \bibinfo {author}
  {\bibfnamefont {T.}~\bibnamefont {Pruttivarasin}}, \bibinfo {author}
  {\bibfnamefont {M.}~\bibnamefont {Hohensee}}, \ and\ \bibinfo {author}
  {\bibfnamefont {H.}~\bibnamefont {H{\"a}ffner}},\ }\bibfield  {title}
  {\enquote {\bibinfo {title} {Strongly enhanced effects of lorentz symmetry
  violation in entangled yb+ ions},}\ }\href@noop {} {\bibfield  {journal}
  {\bibinfo  {journal} {Nature Physics}\ }\textbf {\bibinfo {volume} {12}},\
  \bibinfo {pages} {465--468} (\bibinfo {year} {2016})}\BibitemShut {NoStop}%
\bibitem [{\citenamefont {Wcis{\l}o}\ \emph {et~al.}(2016)\citenamefont
  {Wcis{\l}o}, \citenamefont {Morzy{\'n}ski}, \citenamefont {Bober},
  \citenamefont {Cygan}, \citenamefont {Lisak}, \citenamefont {Ciury{\l}o},\
  and\ \citenamefont {Zawada}}]{wcislo2016experimental}%
  \BibitemOpen
  \bibfield  {author} {\bibinfo {author} {\bibfnamefont {P.}~\bibnamefont
  {Wcis{\l}o}}, \bibinfo {author} {\bibfnamefont {P.}~\bibnamefont
  {Morzy{\'n}ski}}, \bibinfo {author} {\bibfnamefont {M.}~\bibnamefont
  {Bober}}, \bibinfo {author} {\bibfnamefont {A.}~\bibnamefont {Cygan}},
  \bibinfo {author} {\bibfnamefont {D.}~\bibnamefont {Lisak}}, \bibinfo
  {author} {\bibfnamefont {R.}~\bibnamefont {Ciury{\l}o}}, \ and\ \bibinfo
  {author} {\bibfnamefont {M.}~\bibnamefont {Zawada}},\ }\bibfield  {title}
  {\enquote {\bibinfo {title} {Experimental constraint on dark matter detection
  with optical atomic clocks},}\ }\href@noop {} {\bibfield  {journal} {\bibinfo
   {journal} {Nature Astronomy}\ }\textbf {\bibinfo {volume} {1}},\ \bibinfo
  {pages} {0009} (\bibinfo {year} {2016})}\BibitemShut {NoStop}%
\bibitem [{\citenamefont {Safronova}\ \emph {et~al.}(2018)\citenamefont
  {Safronova}, \citenamefont {Budker}, \citenamefont {DeMille}, \citenamefont
  {Kimball}, \citenamefont {Derevianko},\ and\ \citenamefont
  {Clark}}]{safronova2018search}%
  \BibitemOpen
  \bibfield  {author} {\bibinfo {author} {\bibfnamefont {M.}~\bibnamefont
  {Safronova}}, \bibinfo {author} {\bibfnamefont {D.}~\bibnamefont {Budker}},
  \bibinfo {author} {\bibfnamefont {D.}~\bibnamefont {DeMille}}, \bibinfo
  {author} {\bibfnamefont {D.~F.~J.}\ \bibnamefont {Kimball}}, \bibinfo
  {author} {\bibfnamefont {A.}~\bibnamefont {Derevianko}}, \ and\ \bibinfo
  {author} {\bibfnamefont {C.~W.}\ \bibnamefont {Clark}},\ }\bibfield  {title}
  {\enquote {\bibinfo {title} {Search for new physics with atoms and
  molecules},}\ }\href@noop {} {\bibfield  {journal} {\bibinfo  {journal}
  {Reviews of Modern Physics}\ }\textbf {\bibinfo {volume} {90}},\ \bibinfo
  {pages} {025008} (\bibinfo {year} {2018})}\BibitemShut {NoStop}%
\bibitem [{\citenamefont {Bandi}\ \emph {et~al.}(2011)\citenamefont {Bandi},
  \citenamefont {Affolderbach}, \citenamefont {Calosso},\ and\ \citenamefont
  {Mileti}}]{bandi2011high}%
  \BibitemOpen
  \bibfield  {author} {\bibinfo {author} {\bibfnamefont {T.}~\bibnamefont
  {Bandi}}, \bibinfo {author} {\bibfnamefont {C.}~\bibnamefont {Affolderbach}},
  \bibinfo {author} {\bibfnamefont {C.~E.}\ \bibnamefont {Calosso}}, \ and\
  \bibinfo {author} {\bibfnamefont {G.}~\bibnamefont {Mileti}},\ }\bibfield
  {title} {\enquote {\bibinfo {title} {High-performance laser-pumped rubidium
  frequency standard for satellite navigation},}\ }\href@noop {} {\bibfield
  {journal} {\bibinfo  {journal} {Electronics letters}\ }\textbf {\bibinfo
  {volume} {47}},\ \bibinfo {pages} {698--699} (\bibinfo {year}
  {2011})}\BibitemShut {NoStop}%
\bibitem [{\citenamefont {Mallette}, \citenamefont {White},\ and\ \citenamefont
  {Rochat}(2010)}]{mallette2010space}%
  \BibitemOpen
  \bibfield  {author} {\bibinfo {author} {\bibfnamefont {L.~A.}\ \bibnamefont
  {Mallette}}, \bibinfo {author} {\bibfnamefont {J.}~\bibnamefont {White}}, \
  and\ \bibinfo {author} {\bibfnamefont {P.}~\bibnamefont {Rochat}},\
  }\bibfield  {title} {\enquote {\bibinfo {title} {Space qualified frequency
  sources (clocks) for current and future gnss applications},}\ }in\ \href@noop
  {} {\emph {\bibinfo {booktitle} {IEEE/ION Position, Location and Navigation
  Symposium}}}\ (\bibinfo {organization} {IEEE},\ \bibinfo {year} {2010})\ pp.\
  \bibinfo {pages} {903--908}\BibitemShut {NoStop}%
\bibitem [{\citenamefont {Prestage}\ and\ \citenamefont
  {Weaver}(2007)}]{prestage2007atomic}%
  \BibitemOpen
  \bibfield  {author} {\bibinfo {author} {\bibfnamefont {J.~D.}\ \bibnamefont
  {Prestage}}\ and\ \bibinfo {author} {\bibfnamefont {G.~L.}\ \bibnamefont
  {Weaver}},\ }\bibfield  {title} {\enquote {\bibinfo {title} {Atomic clocks
  and oscillators for deep-space navigation and radio science},}\ }\href@noop
  {} {\bibfield  {journal} {\bibinfo  {journal} {Proceedings of the IEEE}\
  }\textbf {\bibinfo {volume} {95}},\ \bibinfo {pages} {2235--2247} (\bibinfo
  {year} {2007})}\BibitemShut {NoStop}%
\bibitem [{\citenamefont {Burt}\ \emph {et~al.}(2021)\citenamefont {Burt},
  \citenamefont {Prestage}, \citenamefont {Tjoelker}, \citenamefont {Enzer},
  \citenamefont {Kuang}, \citenamefont {Murphy}, \citenamefont {Robison},
  \citenamefont {Seubert}, \citenamefont {Wang},\ and\ \citenamefont
  {Ely}}]{burt2021demonstration}%
  \BibitemOpen
  \bibfield  {author} {\bibinfo {author} {\bibfnamefont {E.}~\bibnamefont
  {Burt}}, \bibinfo {author} {\bibfnamefont {J.}~\bibnamefont {Prestage}},
  \bibinfo {author} {\bibfnamefont {R.}~\bibnamefont {Tjoelker}}, \bibinfo
  {author} {\bibfnamefont {D.}~\bibnamefont {Enzer}}, \bibinfo {author}
  {\bibfnamefont {D.}~\bibnamefont {Kuang}}, \bibinfo {author} {\bibfnamefont
  {D.}~\bibnamefont {Murphy}}, \bibinfo {author} {\bibfnamefont
  {D.}~\bibnamefont {Robison}}, \bibinfo {author} {\bibfnamefont
  {J.}~\bibnamefont {Seubert}}, \bibinfo {author} {\bibfnamefont
  {R.}~\bibnamefont {Wang}}, \ and\ \bibinfo {author} {\bibfnamefont
  {T.}~\bibnamefont {Ely}},\ }\bibfield  {title} {\enquote {\bibinfo {title}
  {Demonstration of a trapped-ion atomic clock in space},}\ }\href@noop {}
  {\bibfield  {journal} {\bibinfo  {journal} {Nature}\ }\textbf {\bibinfo
  {volume} {595}},\ \bibinfo {pages} {43--47} (\bibinfo {year}
  {2021})}\BibitemShut {NoStop}%
\bibitem [{\citenamefont {Piester}\ \emph {et~al.}(2011)\citenamefont
  {Piester}, \citenamefont {Rost}, \citenamefont {Fujieda}, \citenamefont
  {Feldmann},\ and\ \citenamefont {Bauch}}]{piester2011remote}%
  \BibitemOpen
  \bibfield  {author} {\bibinfo {author} {\bibfnamefont {D.}~\bibnamefont
  {Piester}}, \bibinfo {author} {\bibfnamefont {M.}~\bibnamefont {Rost}},
  \bibinfo {author} {\bibfnamefont {M.}~\bibnamefont {Fujieda}}, \bibinfo
  {author} {\bibfnamefont {T.}~\bibnamefont {Feldmann}}, \ and\ \bibinfo
  {author} {\bibfnamefont {A.}~\bibnamefont {Bauch}},\ }\bibfield  {title}
  {\enquote {\bibinfo {title} {Remote atomic clock synchronization via
  satellites and optical fibers},}\ }\href@noop {} {\bibfield  {journal}
  {\bibinfo  {journal} {Advances in Radio Science}\ }\textbf {\bibinfo {volume}
  {9}},\ \bibinfo {pages} {1--7} (\bibinfo {year} {2011})}\BibitemShut
  {NoStop}%
\bibitem [{\citenamefont {Diddams}\ \emph {et~al.}(2004)\citenamefont
  {Diddams}, \citenamefont {Bergquist}, \citenamefont {Jefferts},\ and\
  \citenamefont {Oates}}]{diddams2004standards}%
  \BibitemOpen
  \bibfield  {author} {\bibinfo {author} {\bibfnamefont {S.~A.}\ \bibnamefont
  {Diddams}}, \bibinfo {author} {\bibfnamefont {J.~C.}\ \bibnamefont
  {Bergquist}}, \bibinfo {author} {\bibfnamefont {S.~R.}\ \bibnamefont
  {Jefferts}}, \ and\ \bibinfo {author} {\bibfnamefont {C.~W.}\ \bibnamefont
  {Oates}},\ }\bibfield  {title} {\enquote {\bibinfo {title} {Standards of time
  and frequency at the outset of the 21st century},}\ }\href@noop {} {\bibfield
   {journal} {\bibinfo  {journal} {Science}\ }\textbf {\bibinfo {volume}
  {306}},\ \bibinfo {pages} {1318--1324} (\bibinfo {year} {2004})}\BibitemShut
  {NoStop}%
\bibitem [{\citenamefont {Burt}, \citenamefont {Diener},\ and\ \citenamefont
  {Tjoelker}(2008)}]{burt2008compensated}%
  \BibitemOpen
  \bibfield  {author} {\bibinfo {author} {\bibfnamefont {E.~A.}\ \bibnamefont
  {Burt}}, \bibinfo {author} {\bibfnamefont {W.~A.}\ \bibnamefont {Diener}}, \
  and\ \bibinfo {author} {\bibfnamefont {R.~L.}\ \bibnamefont {Tjoelker}},\
  }\bibfield  {title} {\enquote {\bibinfo {title} {A compensated multi-pole
  linear ion trap mercury frequency standard for ultra-stable timekeeping},}\
  }\href@noop {} {\bibfield  {journal} {\bibinfo  {journal} {IEEE transactions
  on ultrasonics, ferroelectrics, and frequency control}\ }\textbf {\bibinfo
  {volume} {55}},\ \bibinfo {pages} {2586--2595} (\bibinfo {year}
  {2008})}\BibitemShut {NoStop}%
\bibitem [{\citenamefont {Yan}\ \emph {et~al.}(2022)\citenamefont {Yan},
  \citenamefont {Liu}, \citenamefont {Chen}, \citenamefont {Liu}, \citenamefont
  {Liu}, \citenamefont {Wang},\ and\ \citenamefont {She}}]{yan2022research}%
  \BibitemOpen
  \bibfield  {author} {\bibinfo {author} {\bibfnamefont {B.~B.}\ \bibnamefont
  {Yan}}, \bibinfo {author} {\bibfnamefont {H.}~\bibnamefont {Liu}}, \bibinfo
  {author} {\bibfnamefont {Y.~H.}\ \bibnamefont {Chen}}, \bibinfo {author}
  {\bibfnamefont {G.}~\bibnamefont {Liu}}, \bibinfo {author} {\bibfnamefont
  {W.~C.}\ \bibnamefont {Liu}}, \bibinfo {author} {\bibfnamefont
  {J.}~\bibnamefont {Wang}}, \ and\ \bibinfo {author} {\bibfnamefont
  {L.}~\bibnamefont {She}},\ }\bibfield  {title} {\enquote {\bibinfo {title}
  {Research progress on mercury ion microwave clock for time keeping},}\ }in\
  \href@noop {} {\emph {\bibinfo {booktitle} {China Satellite Navigation
  Conference (CSNC 2022) Proceedings: Volume III}}}\ (\bibinfo {organization}
  {Springer},\ \bibinfo {year} {2022})\ pp.\ \bibinfo {pages}
  {345--352}\BibitemShut {NoStop}%
\bibitem [{\citenamefont {Mulholland}\ \emph {et~al.}(2019)\citenamefont
  {Mulholland}, \citenamefont {Klein}, \citenamefont {Barwood}, \citenamefont
  {Donnellan}, \citenamefont {Gentle}, \citenamefont {Huang}, \citenamefont
  {Walsh}, \citenamefont {Baird},\ and\ \citenamefont
  {Gill}}]{mulholland2019laser}%
  \BibitemOpen
  \bibfield  {author} {\bibinfo {author} {\bibfnamefont {S.}~\bibnamefont
  {Mulholland}}, \bibinfo {author} {\bibfnamefont {H.}~\bibnamefont {Klein}},
  \bibinfo {author} {\bibfnamefont {G.}~\bibnamefont {Barwood}}, \bibinfo
  {author} {\bibfnamefont {S.}~\bibnamefont {Donnellan}}, \bibinfo {author}
  {\bibfnamefont {D.}~\bibnamefont {Gentle}}, \bibinfo {author} {\bibfnamefont
  {G.}~\bibnamefont {Huang}}, \bibinfo {author} {\bibfnamefont
  {G.}~\bibnamefont {Walsh}}, \bibinfo {author} {\bibfnamefont
  {P.}~\bibnamefont {Baird}}, \ and\ \bibinfo {author} {\bibfnamefont
  {P.}~\bibnamefont {Gill}},\ }\bibfield  {title} {\enquote {\bibinfo {title}
  {Laser-cooled ytterbium-ion microwave frequency standard},}\ }\href@noop {}
  {\bibfield  {journal} {\bibinfo  {journal} {Applied Physics B}\ }\textbf
  {\bibinfo {volume} {125}},\ \bibinfo {pages} {198} (\bibinfo {year}
  {2019})}\BibitemShut {NoStop}%
\bibitem [{\citenamefont {Xin}\ \emph {et~al.}(2022)\citenamefont {Xin},
  \citenamefont {Qin}, \citenamefont {Miao}, \citenamefont {Chen},
  \citenamefont {Zheng}, \citenamefont {Han}, \citenamefont {Zhang},\ and\
  \citenamefont {Wang}}]{xin2022laser}%
  \BibitemOpen
  \bibfield  {author} {\bibinfo {author} {\bibfnamefont {N.~C.}\ \bibnamefont
  {Xin}}, \bibinfo {author} {\bibfnamefont {H.~R.}\ \bibnamefont {Qin}},
  \bibinfo {author} {\bibfnamefont {S.~N.}\ \bibnamefont {Miao}}, \bibinfo
  {author} {\bibfnamefont {Y.~T.}\ \bibnamefont {Chen}}, \bibinfo {author}
  {\bibfnamefont {Y.}~\bibnamefont {Zheng}}, \bibinfo {author} {\bibfnamefont
  {J.~Z.}\ \bibnamefont {Han}}, \bibinfo {author} {\bibfnamefont {J.~W.}\
  \bibnamefont {Zhang}}, \ and\ \bibinfo {author} {\bibfnamefont {L.~J.}\
  \bibnamefont {Wang}},\ }\bibfield  {title} {\enquote {\bibinfo {title}
  {Laser-cooled 171 yb+ microwave frequency standard with a short-term
  frequency instability of $8.5\times 10^{-13}/\sqrt{\tau}$},}\ }\href@noop {}
  {\bibfield  {journal} {\bibinfo  {journal} {Optics Express}\ }\textbf
  {\bibinfo {volume} {30}},\ \bibinfo {pages} {14574--14585} (\bibinfo {year}
  {2022})}\BibitemShut {NoStop}%
\bibitem [{\citenamefont {Zhang}\ \emph {et~al.}(2014)\citenamefont {Zhang},
  \citenamefont {Wang}, \citenamefont {Miao}, \citenamefont {Wang},\ and\
  \citenamefont {Wang}}]{zhang2014toward}%
  \BibitemOpen
  \bibfield  {author} {\bibinfo {author} {\bibfnamefont {J.~W.}\ \bibnamefont
  {Zhang}}, \bibinfo {author} {\bibfnamefont {S.~G.}\ \bibnamefont {Wang}},
  \bibinfo {author} {\bibfnamefont {K.}~\bibnamefont {Miao}}, \bibinfo {author}
  {\bibfnamefont {Z.~B.}\ \bibnamefont {Wang}}, \ and\ \bibinfo {author}
  {\bibfnamefont {L.~J.}\ \bibnamefont {Wang}},\ }\bibfield  {title} {\enquote
  {\bibinfo {title} {Toward a transportable microwave frequency standard based
  on laser-cooled 113 cd+ ions},}\ }\href@noop {} {\bibfield  {journal}
  {\bibinfo  {journal} {Applied Physics B}\ }\textbf {\bibinfo {volume}
  {114}},\ \bibinfo {pages} {183--187} (\bibinfo {year} {2014})}\BibitemShut
  {NoStop}%
\bibitem [{\citenamefont {Miao}\ \emph {et~al.}(2015)\citenamefont {Miao},
  \citenamefont {Zhang}, \citenamefont {Sun}, \citenamefont {Wang},
  \citenamefont {Zhang}, \citenamefont {Liang},\ and\ \citenamefont
  {Wang}}]{miao2015high}%
  \BibitemOpen
  \bibfield  {author} {\bibinfo {author} {\bibfnamefont {K.}~\bibnamefont
  {Miao}}, \bibinfo {author} {\bibfnamefont {J.~W.}\ \bibnamefont {Zhang}},
  \bibinfo {author} {\bibfnamefont {X.~L.}\ \bibnamefont {Sun}}, \bibinfo
  {author} {\bibfnamefont {S.~G.}\ \bibnamefont {Wang}}, \bibinfo {author}
  {\bibfnamefont {A.~M.}\ \bibnamefont {Zhang}}, \bibinfo {author}
  {\bibfnamefont {K.}~\bibnamefont {Liang}}, \ and\ \bibinfo {author}
  {\bibfnamefont {L.~J.}\ \bibnamefont {Wang}},\ }\bibfield  {title} {\enquote
  {\bibinfo {title} {High accuracy measurement of the ground-state hyperfine
  splitting in a 113 cd+ microwave clock},}\ }\href@noop {} {\bibfield
  {journal} {\bibinfo  {journal} {Optics letters}\ }\textbf {\bibinfo {volume}
  {40}},\ \bibinfo {pages} {4249--4252} (\bibinfo {year} {2015})}\BibitemShut
  {NoStop}%
\bibitem [{\citenamefont {Miao}\ \emph
  {et~al.}(2021{\natexlab{a}})\citenamefont {Miao}, \citenamefont {Zhang},
  \citenamefont {Qin}, \citenamefont {Xin}, \citenamefont {Han},\ and\
  \citenamefont {Wang}}]{miao2021precision}%
  \BibitemOpen
  \bibfield  {author} {\bibinfo {author} {\bibfnamefont {S.~N.}\ \bibnamefont
  {Miao}}, \bibinfo {author} {\bibfnamefont {J.~W.}\ \bibnamefont {Zhang}},
  \bibinfo {author} {\bibfnamefont {H.~R.}\ \bibnamefont {Qin}}, \bibinfo
  {author} {\bibfnamefont {N.~C.}\ \bibnamefont {Xin}}, \bibinfo {author}
  {\bibfnamefont {J.~Z.}\ \bibnamefont {Han}}, \ and\ \bibinfo {author}
  {\bibfnamefont {L.~J.}\ \bibnamefont {Wang}},\ }\bibfield  {title} {\enquote
  {\bibinfo {title} {Precision determination of the ground-state hyperfine
  splitting of trapped 113 cd+ ions},}\ }\href@noop {} {\bibfield  {journal}
  {\bibinfo  {journal} {Optics letters}\ }\textbf {\bibinfo {volume} {46}},\
  \bibinfo {pages} {5882--5885} (\bibinfo {year}
  {2021}{\natexlab{a}})}\BibitemShut {NoStop}%
\bibitem [{\citenamefont {Qin}\ \emph {et~al.}(2022)\citenamefont {Qin},
  \citenamefont {Miao}, \citenamefont {Han}, \citenamefont {Xin}, \citenamefont
  {Chen}, \citenamefont {Zhang}, \citenamefont {Wang} \emph
  {et~al.}}]{qin2022high}%
  \BibitemOpen
  \bibfield  {author} {\bibinfo {author} {\bibfnamefont {H.~R.}\ \bibnamefont
  {Qin}}, \bibinfo {author} {\bibfnamefont {S.~N.}\ \bibnamefont {Miao}},
  \bibinfo {author} {\bibfnamefont {J.~Z.}\ \bibnamefont {Han}}, \bibinfo
  {author} {\bibfnamefont {N.~C.}\ \bibnamefont {Xin}}, \bibinfo {author}
  {\bibfnamefont {Y.~T.}\ \bibnamefont {Chen}}, \bibinfo {author}
  {\bibfnamefont {J.~W.}\ \bibnamefont {Zhang}}, \bibinfo {author}
  {\bibfnamefont {L.~J.}\ \bibnamefont {Wang}},  \emph {et~al.},\ }\bibfield
  {title} {\enquote {\bibinfo {title} {High-performance microwave frequency
  standard based on sympathetically cooled ions},}\ }\href@noop {} {\bibfield
  {journal} {\bibinfo  {journal} {Physical Review Applied}\ }\textbf {\bibinfo
  {volume} {18}},\ \bibinfo {pages} {024023} (\bibinfo {year}
  {2022})}\BibitemShut {NoStop}%
\bibitem [{\citenamefont {Zuo}\ \emph {et~al.}(2018)\citenamefont {Zuo},
  \citenamefont {Han}, \citenamefont {Wei}, \citenamefont {Zhang},\ and\
  \citenamefont {Wang}}]{zuo2018progress}%
  \BibitemOpen
  \bibfield  {author} {\bibinfo {author} {\bibfnamefont {Y.~N.}\ \bibnamefont
  {Zuo}}, \bibinfo {author} {\bibfnamefont {J.~Z.}\ \bibnamefont {Han}},
  \bibinfo {author} {\bibfnamefont {L.}~\bibnamefont {Wei}}, \bibinfo {author}
  {\bibfnamefont {J.~W.}\ \bibnamefont {Zhang}}, \ and\ \bibinfo {author}
  {\bibfnamefont {L.~J.}\ \bibnamefont {Wang}},\ }\bibfield  {title} {\enquote
  {\bibinfo {title} {Progress towards a cadimium ion microwave clock based on
  sympathetic cooling},}\ }in\ \href@noop {} {\emph {\bibinfo {booktitle} {2018
  IEEE International Frequency Control Symposium (IFCS)}}}\ (\bibinfo
  {organization} {IEEE},\ \bibinfo {year} {2018})\ pp.\ \bibinfo {pages}
  {1--3}\BibitemShut {NoStop}%
\bibitem [{\citenamefont {Han}\ \emph {et~al.}(2021)\citenamefont {Han},
  \citenamefont {Qin}, \citenamefont {Xin}, \citenamefont {Yu}, \citenamefont
  {Dzuba}, \citenamefont {Zhang},\ and\ \citenamefont {Wang}}]{han2021toward}%
  \BibitemOpen
  \bibfield  {author} {\bibinfo {author} {\bibfnamefont {J.~Z.}\ \bibnamefont
  {Han}}, \bibinfo {author} {\bibfnamefont {H.~R.}\ \bibnamefont {Qin}},
  \bibinfo {author} {\bibfnamefont {N.~C.}\ \bibnamefont {Xin}}, \bibinfo
  {author} {\bibfnamefont {Y.~M.}\ \bibnamefont {Yu}}, \bibinfo {author}
  {\bibfnamefont {V.}~\bibnamefont {Dzuba}}, \bibinfo {author} {\bibfnamefont
  {J.~W.}\ \bibnamefont {Zhang}}, \ and\ \bibinfo {author} {\bibfnamefont
  {L.~J.}\ \bibnamefont {Wang}},\ }\bibfield  {title} {\enquote {\bibinfo
  {title} {Toward a high-performance transportable microwave frequency standard
  based on sympathetically cooled 113cd+ ions},}\ }\href@noop {} {\bibfield
  {journal} {\bibinfo  {journal} {Applied Physics Letters}\ }\textbf {\bibinfo
  {volume} {118}},\ \bibinfo {pages} {101103} (\bibinfo {year}
  {2021})}\BibitemShut {NoStop}%
\bibitem [{\citenamefont {Zuo}\ \emph {et~al.}(2019)\citenamefont {Zuo},
  \citenamefont {Han}, \citenamefont {Zhang},\ and\ \citenamefont
  {Wang}}]{zuo2019direct}%
  \BibitemOpen
  \bibfield  {author} {\bibinfo {author} {\bibfnamefont {Y.~N.}\ \bibnamefont
  {Zuo}}, \bibinfo {author} {\bibfnamefont {J.~Z.}\ \bibnamefont {Han}},
  \bibinfo {author} {\bibfnamefont {J.~W.}\ \bibnamefont {Zhang}}, \ and\
  \bibinfo {author} {\bibfnamefont {L.~J.}\ \bibnamefont {Wang}},\ }\bibfield
  {title} {\enquote {\bibinfo {title} {Direct temperature determination of a
  sympathetically cooled large 113cd+ ion crystal for a microwave clock},}\
  }\href@noop {} {\bibfield  {journal} {\bibinfo  {journal} {Applied Physics
  Letters}\ }\textbf {\bibinfo {volume} {115}},\ \bibinfo {pages} {061103}
  (\bibinfo {year} {2019})}\BibitemShut {NoStop}%
\bibitem [{\citenamefont {Miao}\ \emph {et~al.}(2023)\citenamefont {Miao},
  \citenamefont {Qin}, \citenamefont {Xin}, \citenamefont {Han}, \citenamefont
  {Chen}, \citenamefont {Zhang},\ and\ \citenamefont
  {Wang}}]{miao2023sympathetic}%
  \BibitemOpen
  \bibfield  {author} {\bibinfo {author} {\bibfnamefont {S.~N.}\ \bibnamefont
  {Miao}}, \bibinfo {author} {\bibfnamefont {H.~R.}\ \bibnamefont {Qin}},
  \bibinfo {author} {\bibfnamefont {N.~C.}\ \bibnamefont {Xin}}, \bibinfo
  {author} {\bibfnamefont {J.~Z.}\ \bibnamefont {Han}}, \bibinfo {author}
  {\bibfnamefont {Y.~T.}\ \bibnamefont {Chen}}, \bibinfo {author}
  {\bibfnamefont {J.~W.}\ \bibnamefont {Zhang}}, \ and\ \bibinfo {author}
  {\bibfnamefont {L.~J.}\ \bibnamefont {Wang}},\ }\bibfield  {title} {\enquote
  {\bibinfo {title} {Sympathetic cooling of a large 113cd+ ion crystal with
  40ca+ in a linear paul trap},}\ }\href@noop {} {\bibfield  {journal}
  {\bibinfo  {journal} {Chinese Journal of Physics}\ }\textbf {\bibinfo
  {volume} {83}},\ \bibinfo {pages} {242--252} (\bibinfo {year}
  {2023})}\BibitemShut {NoStop}%
\bibitem [{\citenamefont {Berkeland}\ \emph {et~al.}(1998)\citenamefont
  {Berkeland}, \citenamefont {Miller}, \citenamefont {Bergquist}, \citenamefont
  {Itano},\ and\ \citenamefont {Wineland}}]{berkeland1998minimization}%
  \BibitemOpen
  \bibfield  {author} {\bibinfo {author} {\bibfnamefont {D.}~\bibnamefont
  {Berkeland}}, \bibinfo {author} {\bibfnamefont {J.}~\bibnamefont {Miller}},
  \bibinfo {author} {\bibfnamefont {J.~C.}\ \bibnamefont {Bergquist}}, \bibinfo
  {author} {\bibfnamefont {W.~M.}\ \bibnamefont {Itano}}, \ and\ \bibinfo
  {author} {\bibfnamefont {D.~J.}\ \bibnamefont {Wineland}},\ }\bibfield
  {title} {\enquote {\bibinfo {title} {Minimization of ion micromotion in a
  paul trap},}\ }\href@noop {} {\bibfield  {journal} {\bibinfo  {journal}
  {Journal of applied physics}\ }\textbf {\bibinfo {volume} {83}},\ \bibinfo
  {pages} {5025--5033} (\bibinfo {year} {1998})}\BibitemShut {NoStop}%
\bibitem [{\citenamefont {Miao}\ \emph {et~al.}(2022)\citenamefont {Miao},
  \citenamefont {Zhang}, \citenamefont {Zheng}, \citenamefont {Qin},
  \citenamefont {Xin}, \citenamefont {Chen}, \citenamefont {Han},\ and\
  \citenamefont {Wang}}]{miao2022second}%
  \BibitemOpen
  \bibfield  {author} {\bibinfo {author} {\bibfnamefont {S.~N.}\ \bibnamefont
  {Miao}}, \bibinfo {author} {\bibfnamefont {J.~W.}\ \bibnamefont {Zhang}},
  \bibinfo {author} {\bibfnamefont {Y.}~\bibnamefont {Zheng}}, \bibinfo
  {author} {\bibfnamefont {H.~R.}\ \bibnamefont {Qin}}, \bibinfo {author}
  {\bibfnamefont {N.~C.}\ \bibnamefont {Xin}}, \bibinfo {author} {\bibfnamefont
  {Y.~T.}\ \bibnamefont {Chen}}, \bibinfo {author} {\bibfnamefont {J.~Z.}\
  \bibnamefont {Han}}, \ and\ \bibinfo {author} {\bibfnamefont {L.~J.}\
  \bibnamefont {Wang}},\ }\bibfield  {title} {\enquote {\bibinfo {title}
  {Second-order doppler frequency shifts of trapped ions in a linear paul
  trap},}\ }\href@noop {} {\bibfield  {journal} {\bibinfo  {journal} {Physical
  Review A}\ }\textbf {\bibinfo {volume} {106}},\ \bibinfo {pages} {033121}
  (\bibinfo {year} {2022})}\BibitemShut {NoStop}%
\bibitem [{\citenamefont {Miao}\ \emph
  {et~al.}(2021{\natexlab{b}})\citenamefont {Miao}, \citenamefont {Zhang},
  \citenamefont {Xin}, \citenamefont {Guo}, \citenamefont {Hu}, \citenamefont
  {Shi}, \citenamefont {Qin}, \citenamefont {Han},\ and\ \citenamefont
  {Wang}}]{miao2021research}%
  \BibitemOpen
  \bibfield  {author} {\bibinfo {author} {\bibfnamefont {S.~N.}\ \bibnamefont
  {Miao}}, \bibinfo {author} {\bibfnamefont {J.~W.}\ \bibnamefont {Zhang}},
  \bibinfo {author} {\bibfnamefont {N.~C.}\ \bibnamefont {Xin}}, \bibinfo
  {author} {\bibfnamefont {L.~M.}\ \bibnamefont {Guo}}, \bibinfo {author}
  {\bibfnamefont {H.~X.}\ \bibnamefont {Hu}}, \bibinfo {author} {\bibfnamefont
  {W.~X.}\ \bibnamefont {Shi}}, \bibinfo {author} {\bibfnamefont {H.~R.}\
  \bibnamefont {Qin}}, \bibinfo {author} {\bibfnamefont {J.~Z.}\ \bibnamefont
  {Han}}, \ and\ \bibinfo {author} {\bibfnamefont {L.~J.}\ \bibnamefont
  {Wang}},\ }\bibfield  {title} {\enquote {\bibinfo {title} {Research on
  sympathetic cooling 113 cd+-174 yb+ system by molecular dynamics
  simulation},}\ }in\ \href@noop {} {\emph {\bibinfo {booktitle} {2021 Joint
  Conference of the European Frequency and Time Forum and IEEE International
  Frequency Control Symposium (EFTF/IFCS)}}}\ (\bibinfo {organization} {IEEE},\
  \bibinfo {year} {2021})\ pp.\ \bibinfo {pages} {1--3}\BibitemShut {NoStop}%
\bibitem [{\citenamefont {Guo}(2021)}]{guoliming}%
  \BibitemOpen
  \bibfield  {author} {\bibinfo {author} {\bibfnamefont {L.}~\bibnamefont
  {Guo}},\ }\emph {\bibinfo {title} {Study on Spectrum and Laser Frequency
  Stabilization of Yb Hollow Cathode Lamp}},\ \href@noop {} {Master's thesis},\
  \bibinfo  {school} {Tsinghua University} (\bibinfo {year} {2021})\BibitemShut
  {NoStop}%
\bibitem [{\citenamefont {Hornek{\ae}r}(2000)}]{hornekaer2000single}%
  \BibitemOpen
  \bibfield  {author} {\bibinfo {author} {\bibfnamefont {L.}~\bibnamefont
  {Hornek{\ae}r}},\ }\emph {\bibinfo {title} {Single-and multi-species Coulomb
  ion crystals: Structures, dynamics and sympathetic cooling}},\ \href@noop {}
  {Ph.D. thesis},\ \bibinfo  {school} {The University of Aarhus} (\bibinfo
  {year} {2000})\BibitemShut {NoStop}%
\bibitem [{\citenamefont {Hornek{\ae}r}\ \emph {et~al.}(2001)\citenamefont
  {Hornek{\ae}r}, \citenamefont {Kj{\ae}rgaard}, \citenamefont {Thommesen},\
  and\ \citenamefont {Drewsen}}]{hornekaer2001structural}%
  \BibitemOpen
  \bibfield  {author} {\bibinfo {author} {\bibfnamefont {L.}~\bibnamefont
  {Hornek{\ae}r}}, \bibinfo {author} {\bibfnamefont {N.}~\bibnamefont
  {Kj{\ae}rgaard}}, \bibinfo {author} {\bibfnamefont {A.}~\bibnamefont
  {Thommesen}}, \ and\ \bibinfo {author} {\bibfnamefont {M.}~\bibnamefont
  {Drewsen}},\ }\bibfield  {title} {\enquote {\bibinfo {title} {Structural
  properties of two-component coulomb crystals in linear paul traps},}\
  }\href@noop {} {\bibfield  {journal} {\bibinfo  {journal} {Physical review
  letters}\ }\textbf {\bibinfo {volume} {86}},\ \bibinfo {pages} {1994}
  (\bibinfo {year} {2001})}\BibitemShut {NoStop}%
\bibitem [{\citenamefont {O’Neil}(1981)}]{o1981centrifugal}%
  \BibitemOpen
  \bibfield  {author} {\bibinfo {author} {\bibfnamefont {T.}~\bibnamefont
  {O’Neil}},\ }\bibfield  {title} {\enquote {\bibinfo {title} {Centrifugal
  separation of a multispecies pure ion plasma},}\ }\href@noop {} {\bibfield
  {journal} {\bibinfo  {journal} {The Physics of Fluids}\ }\textbf {\bibinfo
  {volume} {24}},\ \bibinfo {pages} {1447--1451} (\bibinfo {year}
  {1981})}\BibitemShut {NoStop}%
\bibitem [{\citenamefont {Wineland}(1987)}]{wineland1987ion}%
  \BibitemOpen
  \bibfield  {author} {\bibinfo {author} {\bibfnamefont {D.}~\bibnamefont
  {Wineland}},\ }\bibfield  {title} {\enquote {\bibinfo {title} {Ion traps for
  large storage capacity},}\ }in\ \href@noop {} {\emph {\bibinfo {booktitle}
  {Proceedings of the Cooling, Condensation, and Storage of Hydrogen Cluster
  Ions Workshop, Menlo Park}}}\ (\bibinfo {organization} {Citeseer},\ \bibinfo
  {year} {1987})\ p.\ \bibinfo {pages} {181}\BibitemShut {NoStop}%
\bibitem [{\citenamefont {Larson}\ \emph {et~al.}(1986)\citenamefont {Larson},
  \citenamefont {Bergquist}, \citenamefont {Bollinger}, \citenamefont {Itano},\
  and\ \citenamefont {Wineland}}]{larson1986sympathetic}%
  \BibitemOpen
  \bibfield  {author} {\bibinfo {author} {\bibfnamefont {D.}~\bibnamefont
  {Larson}}, \bibinfo {author} {\bibfnamefont {J.~C.}\ \bibnamefont
  {Bergquist}}, \bibinfo {author} {\bibfnamefont {J.~J.}\ \bibnamefont
  {Bollinger}}, \bibinfo {author} {\bibfnamefont {W.~M.}\ \bibnamefont
  {Itano}}, \ and\ \bibinfo {author} {\bibfnamefont {D.~J.}\ \bibnamefont
  {Wineland}},\ }\bibfield  {title} {\enquote {\bibinfo {title} {Sympathetic
  cooling of trapped ions: A laser-cooled two-species nonneutral ion plasma},}\
  }\href@noop {} {\bibfield  {journal} {\bibinfo  {journal} {Physical review
  letters}\ }\textbf {\bibinfo {volume} {57}},\ \bibinfo {pages} {70} (\bibinfo
  {year} {1986})}\BibitemShut {NoStop}%
\bibitem [{\citenamefont {Bollinger}\ and\ \citenamefont
  {Wineland}(1984)}]{bollinger1984strongly}%
  \BibitemOpen
  \bibfield  {author} {\bibinfo {author} {\bibfnamefont {J.}~\bibnamefont
  {Bollinger}}\ and\ \bibinfo {author} {\bibfnamefont {D.}~\bibnamefont
  {Wineland}},\ }\bibfield  {title} {\enquote {\bibinfo {title} {Strongly
  coupled nonneutral ion plasma},}\ }\href@noop {} {\bibfield  {journal}
  {\bibinfo  {journal} {Physical review letters}\ }\textbf {\bibinfo {volume}
  {53}},\ \bibinfo {pages} {348} (\bibinfo {year} {1984})}\BibitemShut
  {NoStop}%
\bibitem [{\citenamefont {Han}\ \emph {et~al.}(2022)\citenamefont {Han},
  \citenamefont {Si}, \citenamefont {Qin}, \citenamefont {Xin}, \citenamefont
  {Chen}, \citenamefont {Miao}, \citenamefont {Chen}, \citenamefont {Zhang},\
  and\ \citenamefont {Wang}}]{han2022determination}%
  \BibitemOpen
  \bibfield  {author} {\bibinfo {author} {\bibfnamefont {J.~Z.}\ \bibnamefont
  {Han}}, \bibinfo {author} {\bibfnamefont {R.}~\bibnamefont {Si}}, \bibinfo
  {author} {\bibfnamefont {H.~R.}\ \bibnamefont {Qin}}, \bibinfo {author}
  {\bibfnamefont {N.~C.}\ \bibnamefont {Xin}}, \bibinfo {author} {\bibfnamefont
  {Y.~T.}\ \bibnamefont {Chen}}, \bibinfo {author} {\bibfnamefont {S.~N.}\
  \bibnamefont {Miao}}, \bibinfo {author} {\bibfnamefont {C.~Y.}\ \bibnamefont
  {Chen}}, \bibinfo {author} {\bibfnamefont {J.~W.}\ \bibnamefont {Zhang}}, \
  and\ \bibinfo {author} {\bibfnamefont {L.~J.}\ \bibnamefont {Wang}},\
  }\bibfield  {title} {\enquote {\bibinfo {title} {Determination of hyperfine
  splittings and land{\'e} g j factors of 5 s 2 s 1/2 and 5 p 2 p 1/2, 3/2
  states of cd+ 111, 113 for microwave frequency standards},}\ }\href@noop {}
  {\bibfield  {journal} {\bibinfo  {journal} {Physical Review A}\ }\textbf
  {\bibinfo {volume} {106}},\ \bibinfo {pages} {012821} (\bibinfo {year}
  {2022})}\BibitemShut {NoStop}%
\bibitem [{\citenamefont {Spence}\ and\ \citenamefont
  {McDermott}(1972)}]{spence1972optical}%
  \BibitemOpen
  \bibfield  {author} {\bibinfo {author} {\bibfnamefont {P.}~\bibnamefont
  {Spence}}\ and\ \bibinfo {author} {\bibfnamefont {M.}~\bibnamefont
  {McDermott}},\ }\bibfield  {title} {\enquote {\bibinfo {title} {Optical
  orientation of 6.7 h107 cd},}\ }\href@noop {} {\bibfield  {journal} {\bibinfo
   {journal} {Physics Letters A}\ }\textbf {\bibinfo {volume} {42}},\ \bibinfo
  {pages} {273--274} (\bibinfo {year} {1972})}\BibitemShut {NoStop}%
\bibitem [{\citenamefont {Richerme}(2016)}]{richerme20162d}%
  \BibitemOpen
  \bibfield  {author} {\bibinfo {author} {\bibfnamefont {P.}~\bibnamefont
  {Richerme}},\ }\bibfield  {title} {\enquote {\bibinfo {title} {2d ion
  crystals in radiofrequency traps for quantum simulation},}\ }\href@noop {}
  {\bibfield  {journal} {\bibinfo  {journal} {arXiv preprint arXiv:1604.08523}\
  } (\bibinfo {year} {2016})}\BibitemShut {NoStop}%
\bibitem [{\citenamefont {D’Onofrio}\ \emph {et~al.}(2021)\citenamefont
  {D’Onofrio}, \citenamefont {Xie}, \citenamefont {Rasmusson}, \citenamefont
  {Wolanski}, \citenamefont {Cui},\ and\ \citenamefont
  {Richerme}}]{d2021radial}%
  \BibitemOpen
  \bibfield  {author} {\bibinfo {author} {\bibfnamefont {M.}~\bibnamefont
  {D’Onofrio}}, \bibinfo {author} {\bibfnamefont {Y.}~\bibnamefont {Xie}},
  \bibinfo {author} {\bibfnamefont {A.}~\bibnamefont {Rasmusson}}, \bibinfo
  {author} {\bibfnamefont {E.}~\bibnamefont {Wolanski}}, \bibinfo {author}
  {\bibfnamefont {J.}~\bibnamefont {Cui}}, \ and\ \bibinfo {author}
  {\bibfnamefont {P.}~\bibnamefont {Richerme}},\ }\bibfield  {title} {\enquote
  {\bibinfo {title} {Radial two-dimensional ion crystals in a linear paul
  trap},}\ }\href@noop {} {\bibfield  {journal} {\bibinfo  {journal} {Physical
  Review Letters}\ }\textbf {\bibinfo {volume} {127}},\ \bibinfo {pages}
  {020503} (\bibinfo {year} {2021})}\BibitemShut {NoStop}%
\bibitem [{\citenamefont {Gebert}\ \emph {et~al.}(2015)\citenamefont {Gebert},
  \citenamefont {Wan}, \citenamefont {Wolf}, \citenamefont {Angstmann},
  \citenamefont {Berengut},\ and\ \citenamefont
  {Schmidt}}]{gebert2015precision}%
  \BibitemOpen
  \bibfield  {author} {\bibinfo {author} {\bibfnamefont {F.}~\bibnamefont
  {Gebert}}, \bibinfo {author} {\bibfnamefont {Y.}~\bibnamefont {Wan}},
  \bibinfo {author} {\bibfnamefont {F.}~\bibnamefont {Wolf}}, \bibinfo {author}
  {\bibfnamefont {C.~N.}\ \bibnamefont {Angstmann}}, \bibinfo {author}
  {\bibfnamefont {J.~C.}\ \bibnamefont {Berengut}}, \ and\ \bibinfo {author}
  {\bibfnamefont {P.~O.}\ \bibnamefont {Schmidt}},\ }\bibfield  {title}
  {\enquote {\bibinfo {title} {Precision isotope shift measurements in calcium
  ions using quantum logic detection schemes},}\ }\href@noop {} {\bibfield
  {journal} {\bibinfo  {journal} {Physical review letters}\ }\textbf {\bibinfo
  {volume} {115}},\ \bibinfo {pages} {053003} (\bibinfo {year}
  {2015})}\BibitemShut {NoStop}%
\bibitem [{\citenamefont {Guggemos}\ \emph {et~al.}(2015)\citenamefont
  {Guggemos}, \citenamefont {Heinrich}, \citenamefont {Herrera-Sancho},
  \citenamefont {Blatt},\ and\ \citenamefont {Roos}}]{guggemos2015sympathetic}%
  \BibitemOpen
  \bibfield  {author} {\bibinfo {author} {\bibfnamefont {M.}~\bibnamefont
  {Guggemos}}, \bibinfo {author} {\bibfnamefont {D.}~\bibnamefont {Heinrich}},
  \bibinfo {author} {\bibfnamefont {O.}~\bibnamefont {Herrera-Sancho}},
  \bibinfo {author} {\bibfnamefont {R.}~\bibnamefont {Blatt}}, \ and\ \bibinfo
  {author} {\bibfnamefont {C.}~\bibnamefont {Roos}},\ }\bibfield  {title}
  {\enquote {\bibinfo {title} {Sympathetic cooling and detection of a hot
  trapped ion by a cold one},}\ }\href@noop {} {\bibfield  {journal} {\bibinfo
  {journal} {New Journal of Physics}\ }\textbf {\bibinfo {volume} {17}},\
  \bibinfo {pages} {103001} (\bibinfo {year} {2015})}\BibitemShut {NoStop}%
\bibitem [{\citenamefont {Groot-Berning}\ \emph {et~al.}(2019)\citenamefont
  {Groot-Berning}, \citenamefont {Stopp}, \citenamefont {Jacob}, \citenamefont
  {Budker}, \citenamefont {Haas}, \citenamefont {Renisch}, \citenamefont
  {Runke}, \citenamefont {Th{\"o}rle-Pospiech}, \citenamefont {D{\"u}llmann},\
  and\ \citenamefont {Schmidt-Kaler}}]{groot2019trapping}%
  \BibitemOpen
  \bibfield  {author} {\bibinfo {author} {\bibfnamefont {K.}~\bibnamefont
  {Groot-Berning}}, \bibinfo {author} {\bibfnamefont {F.}~\bibnamefont
  {Stopp}}, \bibinfo {author} {\bibfnamefont {G.}~\bibnamefont {Jacob}},
  \bibinfo {author} {\bibfnamefont {D.}~\bibnamefont {Budker}}, \bibinfo
  {author} {\bibfnamefont {R.}~\bibnamefont {Haas}}, \bibinfo {author}
  {\bibfnamefont {D.}~\bibnamefont {Renisch}}, \bibinfo {author} {\bibfnamefont
  {J.}~\bibnamefont {Runke}}, \bibinfo {author} {\bibfnamefont
  {P.}~\bibnamefont {Th{\"o}rle-Pospiech}}, \bibinfo {author} {\bibfnamefont
  {C.~E.}\ \bibnamefont {D{\"u}llmann}}, \ and\ \bibinfo {author}
  {\bibfnamefont {F.}~\bibnamefont {Schmidt-Kaler}},\ }\bibfield  {title}
  {\enquote {\bibinfo {title} {Trapping and sympathetic cooling of single
  thorium ions for spectroscopy},}\ }\href@noop {} {\bibfield  {journal}
  {\bibinfo  {journal} {Physical Review A}\ }\textbf {\bibinfo {volume} {99}},\
  \bibinfo {pages} {023420} (\bibinfo {year} {2019})}\BibitemShut {NoStop}%
\bibitem [{\citenamefont {Raghunandan}\ \emph {et~al.}(2020)\citenamefont
  {Raghunandan}, \citenamefont {Wolf}, \citenamefont {Ospelkaus}, \citenamefont
  {Schmidt},\ and\ \citenamefont {Weimer}}]{raghunandan2020initialization}%
  \BibitemOpen
  \bibfield  {author} {\bibinfo {author} {\bibfnamefont {M.}~\bibnamefont
  {Raghunandan}}, \bibinfo {author} {\bibfnamefont {F.}~\bibnamefont {Wolf}},
  \bibinfo {author} {\bibfnamefont {C.}~\bibnamefont {Ospelkaus}}, \bibinfo
  {author} {\bibfnamefont {P.~O.}\ \bibnamefont {Schmidt}}, \ and\ \bibinfo
  {author} {\bibfnamefont {H.}~\bibnamefont {Weimer}},\ }\bibfield  {title}
  {\enquote {\bibinfo {title} {Initialization of quantum simulators by
  sympathetic cooling},}\ }\href@noop {} {\bibfield  {journal} {\bibinfo
  {journal} {Science Advances}\ }\textbf {\bibinfo {volume} {6}},\ \bibinfo
  {pages} {eaaw9268} (\bibinfo {year} {2020})}\BibitemShut {NoStop}%
\end{thebibliography}%

\end{document}